\magnification=1200

\centerline {\bf  On the Large $N$ Limit, $W_\infty $ Strings , Star Products , AdS/CFT duality }
\centerline{\bf Nonlinear Sigma  Models  on $AdS$ Spaces and Chern-Simons $p$-Branes  } 
\bigskip
\centerline { Carlos Castro}
\centerline { Center for Theoretical Studies of Physical Systems}
\centerline { Clark Atlanta University, Atlanta, GA. 30314, USA}
\centerline { e-mail : castro@ctsps.cau.edu}
\bigskip
\centerline { June , 2001}
\bigskip

\centerline{\bf ABSTRACT}
\bigskip

It is shown that the large $N$ limit of $SU(N)$ YM in $curved$  $m$-dim backgrounds 
can be subsumed by a higher $m+n$ dimensional gravitational theory which can be identified 
to an $m$-dim generally invariant gauge theory of diffs $N$, 
where $N$ is an $n$-dim internal space ( Cho, Sho, Park, Yoon ). 
Based on these findings, a very plausible 
geometrical interpretation of the $AdS/CFT$ correspondence could be given. 
Conformally invariant sigma models in $D=2n$ dimensions with target non-compact
$SO(2n,1)$ groups are reviewed. Despite the non-compact
nature of the $SO(2n,1)$ , the classical action and Hamiltonian are positive
definite. Instanton field configurations are found to correspond geometrically to conformal 
``stereographic''  mappings of $R^{2n}$ into the Euclidean signature $AdS_{2n}$ spaces. 
The relation between Self Dual branes and Chern-Simons branes , High Dimensional Knots ,  
follows. 
A detailed discussion on $W_\infty $ symmetry is given and we outline 
the Vasiliev procedure to construct an action involving  higher spin massless fields in $AdS_4$. 
This $AdS_4$ spacetime higher spin theory  should have a one-to-one correspondence 
to noncritical $W_\infty$ strings propagating on $AdS_4 \times S^7$.

\bigskip

\centerline{\bf I. INTRODUCTION    }
\bigskip

Conformal invariance in physics has had a tremendous success. One of the main reasons why conformal symmetry in 
$2D$ CFT has been so powerful is because the conformal group in two dimensions is infinite dimensional which allows to solve the theory exactly. 
CFT has had numerous applications in string theory, critical phenomena in statistical mechanics, $2D$ Gravity, Random Matrix models, 
$2D$ Topological QFT,...Imposing conformal invariance at the quantum level was instrumental in constraining the types of curved backgrounds on 
which strings can propagate ; i.e they must be solutions of the Einstein-Yang-Mills equations. 
Extensions of CFT based on the Virasoro algebra spawned the development of $W_\infty$ symmetries , 
that are higher conformal spin extensions of the Virasoro algebra., $ s = 2,3, 4, ...\infty$. 
Gauging such $W_\infty$ symmetries furnished $W_\infty$ gravity and allowed the construction of $W_\infty $-strings ; 
i.e higher spin fields in the target spacetime [ 14 ].    
 
Conformally invariant  $SO(2n+1)~\sigma$ models in $2n$ dimensions were
of crucial importance in the construction of conformally invariant
Lagrangians, with vanishing world-volume cosmological constant,  for bosonic $p$-branes ( $p+1=2n$ ) [1,2]. 
In particular, Self Dual unit charge instanton solutions were found which correspond to conformal ( stereographic) maps from $R^{2n}\rightarrow S^{2n}$. These models [1] were the higher-dimensional extension of the $SO(5)~\sigma$ 
models of Feisager and Leinaas in $4D$ [3].   
When $p+1=odd$, the authors [2] also built Lagrangians for bosonic extendons ( branes) , however, conformal invariance was lost. 
These sort of Skyrme-like Lagrangians ( quartic derivatives )  allowed the author to construct a $polynomial$ action for 
the spinning membrane after a 
Weyl covariantization process   of the Dolan-Tchrakian action was performed [4].  
The supersymmetric action was invariant under a $modified$  $Q$ supersymmetry transformation 
, a  $Q+S+K$ sum rule, and invariant also under the $homothecy$ transformations . 
A different type of a Weyl invariant spinning membrane action was constructd by [5] at the price of  being 
$non-polynomial$ in the fields, complicating the quantization program.   
The relevance of these quartic derivatives Skyrme type of actions for the spinning membrane has been
recently pointed out by Cho et al [69 ], in a different context,  
as a candidate effective action for $SU(2)$ QCD, based on the Faddeev-Niemi conjecture.

Conformal Field Theories have  risen to more prominence recently mainly due to Maldacena's conjecture on the 
AdS/CFT duality between M/string theory  on    
$AdS_d\times S^{D-d}$ backgrounds and CFT's living on the projective boundary of the $AdS_d$ spaces [6]. Relevant $\sigma$ models with target spaces on certain supergroups have been used to decribe 
$CFT$ on $AdS$ backgrounds with Ramond-Ramond (RR) 
Flux [7]. In particular, various exact $2D~CFT$ on $AdS_{2n+1}$ backgrounds have been constructed 
recently that could be used in building superstring theories propagating on $AdS_{2n+1}$ backgrounds [8]. These $\sigma$ models were based on the standard $SL(2,R)$ WZNW model; i.e 
$\sigma$ models on group manifolds with WZNW terms. 

Conventional $\sigma$-models are based on compact groups like $SO(2n),
SU(n)...$. In general, compact simple groups are used mainly to
simplify the quantization program; i.e the 
quantization is not riddled with the standard problems of 
ghosts  due to the non-positive definite inner products; uniqueness
of  the WZ functional; solvability of the model; positivity of the
Hamiltonian [29,30]. The authors [29] studied $SO(N,1),SU(N,1)$ non-compact sigma models
in two-dimensions and have shown that a dynamical mass generation with asymptotic
freedom is possible and that a sensible unitary quantization program
is  possible by recurring to  a selection
rule in the Hilbert space of states ( only positive definite
norms are allowed).

Setting aside these technical subtleties on the compact simple character of the group,  and motivated by the recent findings on Anti de Sitter spaces,  
we will study conformally invariant 
( noncompact ) $SO(2n-1,2)~\sigma$-models in $2n$ even dimensions; i.e maps from  
$R^{2n}$ to $ SO(2n-1,2)$ and $SO(2n,1)$. Instanton solutions are found
for the latter case, corresponding 
to conformal, ``stereographic `` maps from $R^{2n}$ to $ SO(2n,1)$. The $SO(2n,1)/SO(2n) $ homogeneous group manifold is topologically the
Euclidean signature $AdS_{2n}$ space whose natural isometry group is $SO(2n,1)$.

The  Lorentz signature $AdS_{2n}$ space of constant negative scalar curvature ( negative cosmological constant ) can be viewed 
as the $2n$-dim hypersurface embedded in a pseudo-Euclidean $2n+1$-dim manifold
with coordinates $y^A=y^0, y^1,...,y^{2n}$ and a diagonal metric of signature $ ( d-2, 2  )$ given
by  $\eta_{AB}=diag~(-,+,+,..,+,-)$ with length squared $y^Ay_A$ preserved by the isometry group $SO(2n-1,2)$.  
The  $AdS_{2n}$ is  defined as the geometrical locus  :

$$y^Ay^B\eta_{AB} =-R^2 . \eqno (1a)$$
and has the topology $ S^1 ( time ) \times R^{ d-1} (space ) $. The two timelike directions compactify into a circle $ S^1$ 
giving closed timelike geodesics. 
Whereas de Sitter spaces in $d$-dimensions requires the $(d-1, 1  ) $ signature instead : 
$\eta_{AB} = diag~ (-, +, +, ....+ ) $ and also a change in the sign in the r.h.s.of eq-(1) : 

$$y^Ay^B\eta_{AB} = R^2 . \eqno (1b)$$
and have for topology $ R^1 ( time ) \times S^{ d-1} (space ) $ , a $pseudosphere$   of positive constant scalar curvature 
( positive cosmological constant ). 
Signature subtleties will be of crucial importance in the construction of instanton solutions. The ordinary Hodge dual star operation is $signature$ dependent. 
For example, the double Hodge dual star operation acting on a rank $p$ differential form in a $4D$ space satisfies :

$$^{**}F=s(-1)^{p(4-p)}F.~~s=+1 ~for~(4,0); (2,2).~s=-1~for~(3,1); (1,3)\eqno (2)$$
where we have displayed the explicit signature dependence in the values of $s$. Hence, for a rank two form $F$, in Euclidean $4D$ space and with signature  
$(2,2)$ ( Atiyah-Ward spaces) one can find solutions to the (anti) self dual YM equations : $ ^* F=\pm F$. 
There are no ( real valued solutions ) YM instantons in $4D$ Minkowski space since in the latter one has imaginary eigenvalues  : $ ^* F=\pm  i F$.

The $two~temporal$ variables are required in the embedding process of
$AdS_{2n}$ into the pseudo-Euclidean space  $R^{2n-1,2}$.   We will
show, in fact, 
that it is the noncompact  $SO(2n,1)~\sigma$-model, instead of the 
$SO(2n-1,2)~\sigma$-model , 
that has instanton solutions obeying a double self duality condition similar to the one obeyed by the BPST instanton [9].
It was shown [1] that the Euclidean ( compact) $SO(2n+1)~\sigma$ model instanton solutions are directly related to instanton solutions of 
$SO(2n)$ Generalized YM (GYM) theories in $R^{2n}$ [10] obeying the forementioned double self duality condition.  

In view of the conformally invariant $\sigma$ models/GYM connection in
higher dimensions, 
the next step is to study $p$-branes  ( with $p+1=2n$ ) 
propagating on flat  backgrounds. In particular, the critical case
when the dimensionality of the target space is saturated , $D=p+1=2n$.
Dolan-Tchrakian [1,2] constructed the corresponding conformally
invariant actions with vanishing world-volume
cosmological constant , based on these conformally invariant $\sigma$ models in $2n$-dimensions. 
Upon the algebraic elimination of the auxiliary world-volume metric, Dolan and Tchrakian have shown that one recovers the Dirac-Nambu-Goto action :

$$S=T\int d^{2n} \sigma~\sqrt { |det~G_{\alpha \beta}| }. 
~~~G_{\alpha\beta} =\eta_{\mu\nu}\partial_\alpha X^\mu \partial_\beta X^\nu. \eqno (3)$$
where $G_{\alpha \beta} $ is the induced world-volume metric resulting
from the embedding of the $p+1=2n$ hypervolume into the $D$-dim
target spacetime. 
When the spacetime
dimension is saturated : $D=p+1=2n$ the square root of the
Dirac-Nambu-Goto action simplifies and one obtains the usual Jacobian
for the change of variables from $\sigma $ to $X$.  In such case the Nambu-Goto action is topological  :   
there are no $bulk$ physical local transverse degrees of freedom . 
Such  topological  actions have been studied by  Zaikov [11].

$$ S= T \int d^{2n} \sigma~ 
\partial_{\sigma^1} X^{\mu_1}\wedge ....... \wedge \partial_{\sigma^{2n} } X^{\mu_{2n}}. \eqno (4)$$
where $T$ is the $p$-brane ( extendon) tension. 
For a $p$-brane  whose world-volume has a natural boundary , an integration ( Gauss law) yields :

$$ S_{CS}= T \int_{\partial V} d^{2n-1} \zeta~ 
X^{\mu_1}\wedge \partial_{\zeta^1} X^{\mu_2}\wedge ....... \wedge \partial_{\zeta^{2n-1} } X^{\mu_{2n}}. \eqno (5)$$
one then recovers the action for the Chern-Simons $p'$ brane whose $p'+1$ world-volume variables , 
$\zeta^a,~a=1,2,.....p'+1$, are  integrated over the 
$2n-1$-dim  boundary $\partial V$ of the $2n$-dim domain $V$
associated with the world volume of the open $p$-brane. The value of
$p'$ must be such that $p'+1=p=2n-1$  [11]. Zaikov concluded that
these topological Chern Simons $p'$-branes exit only in target
spacetimes of dimensionality $D=p'+2$. To ensure translational invariance, 
the variables $ X^\mu$ must be understood as $ X^\mu - X^\mu ( 0 )$.

In particular, when the dimensionality of the target spacetime is
saturated , $D=p+1$, one can construct, in addition , self-dual
$p$-brane (extendon) solutions obeying the equations of motion and
constraints ( resulting from $p+1$ reparametrization invariance of the
world-volume) that are directly related to these topological
Chern-Simons $p'$-branes. This holds provided $p'+1=p$ and the embedding manifold is Euclidean [12]. Furthermore, 
when $D=p+1=2n$ one has conformal invariance a well [1,2]. It is in
this fashion how the relationship between the self dual $p$-branes and
Chern Simons $p'=p-1$ branes emerges . 

This is roughly the analogy with Witten's discovery of the one-to-one relationship between 
$3D$ Nonabelian Chern-Simons theories and $2D$ rational CFT. [13]. 
Chern-Simons $p'$-branes have codimension $two$ and for this reason they are the 
Higher Dimensional extensions of Knots ( embeddings of loops $S^1$  into three-dimensions ).   
In a  recent book by Ranicki [47] , the deep connections between High Dimensional Knots , Codimension $q$ surgery and 
algebraic {\bf K }and {\bf L}  Theory are expounded in full mathematical rigour.  
Because Chern-Simons branes have codimension $2$ , one can have two different CS branes living in two complementary dimensions , $d_1, d_2$, 
such that 
$ (p_1 +2) + ( p_2 + 2 ) = p_1+p_2 +4 = d_1 + d_2 = D $ . The latter relation is exactly the same one between a 
$p_1$-brane and its EM dual $p_2$-brane living in $ D $ dimensions.  In this sense, these two CS branes ( high dim knots ) 
intersect $transversely$ from the $D = d_1 + d_2 $ perspective and can be seen as EM duals of each-other.

The contents of this work are the following :

In section {\bf 2} we will show that the Euclidean signature $AdS_{2n}$ space  is an 
instanton solution of the $SO(2n, 1)$ nonlinear conformally invariant 
 $\sigma$ models defined in $R^{2n}$.  
In section {\bf 3.1 } we will establish the relation between self dual  
$p$-branes and Chern-Simons $p'$-branes  ( $ p = p+1 $ ) 
defined in $D$-dim ( pseudo) Euclidean backgrounds when $ D = p+1 = p'+2 $.   
The correspondence between Self Dual branes and Chern-Simons ones is not one-to-one. 
Examples of Chern-Simons branes that are not related to self dual branes are presented. 
This occurs when the world volume of a boundary 
is not necessarily the boundary of a world  volume. 
In {\bf 3.2} we will review how spacetime filling $p$-branes can be obtained directly from a Moyal deformation quantization 
of the Lie algebraic structure of Generalized Yang Mills theories [10] . 
We also discuss actions with a Topological $\theta$ term and show that the large $N$ limit 
of quenched reduced {\bf QCD} in $D=4$ admits hadronic bags and Chern-Simons membranes with $nontrivial$  
boundary dynamics for extended  excitations. 
Using the Zarisksi star product associated with deformations of Nambu-Poisson brackets [63 ] 
allows to see that similiar Chern-Simons branes actions can be obtained from the large 
$N$ limit of $SU(N)$ Matrix Chern-Simons models. Matrix Topological $BF$ models [62] in the large 
$N$ limit  yield Kalb-Ramond couplings to $p$-branes.   

In section {\bf 4.1 } we show how 
an action inspired from a  $BF$ and Chern-Simons model based on the $AdS_4$ isometry group $SO(3, 2)$, 
with the inclusion of a Higgs potential term, 
furnishes the MacDowell-Mansouri-Chamseddine-West  action for gravity with a cosmological constant 
and the Gauss-Bonnet Topological invariant . The $AdS_4$ space is a vacuum solution of this model. 
In section {\bf 4.2 } we study the Lagrangians displaying new couplings of matter and gravity to 
$SU(N)$ Yang Mills theory and arrive to a Chern-Simons-like  action for a $dualized$  
$SU(N)$ Yang-Mills ${\tilde G}_{\mu\nu}$ 
field.  In the case of $AdS_4$ , the  correspondence ( not an identity ) to  a Chern-Simons  membrane living on the boundary  
$S^1 \times S^2$ , and the large $N$ limit of this $dualized$  $SU(N)$ Chern-Simons-like action is pointed out.

In {\bf 5} we show that there are ambiguities 
in taking  the $N = \infty$ limit of $SU(N)$ YM in $curved$ backgrounds , by using the naive quenched-reduced approximation, 
that generated the  the Dolan-Tchrakian action, in the conformal gauge, for  $3$-branes  moving in a flat target spacetime backgrounds .  
The correct large $N$ limit of $SU(N)$ YM in $curved$ backgrounds requires the Fedosov product and trace [ 18 ] and is shown to be related  
to a generally invariant $m$-dimensional gauge theory of ( volume-preserving )  diffs of an internal $n$-dim manifold. 
Such theory   can be identified to a ( restricted ) Einstein gravitational theory in $m+n$-dimensions [ 37] , 
and in this particular case  : $ m = n$ , corresponds  to the $cotangent$ space ( bundle ) of an $m$-dim symplectic manifold.   
A particular example of gauge theories based on infinite dimensional algebras  are the gauge theories of 
the Virasoro algebra ( diffs of $S^1$) [38] ; the $w_\infty $ algebra of symplectic diffs in two-dimensions ( $R^2$) [39] 
and the $w_\infty^\infty$ algebras, the infinite colour limit of $w_\infty$ [ 61] ,  of symplectic diffs in four dimensions.   
Based on these findings, a very plausible geometrical interpretation of the Maldacena $AdS/CFT$ duality could be given.  
This is presented at the 
end of this section.

Finally in the concluding section a detailed discussion is presented on the role that 
$W_\infty$ algebras and $W_\infty$ strings have in the $AdS/CFT$ correspondence. 
The star product of Vasiliev's construction of 
higher spin gauge theories $ s = 2, 3, 4, ....\infty$ in curved $AdS$ backgrounds is used to describe  a 
deformation of a BF-CS-Higss action of {\bf 4.1}. Such Vasiliev deformed action governs 
the nonlinear and nonlocal dynamics of the higher spin fields in $ AdS_4$ spacetime.    
We argue why the latter theory may have a one-to-one correspondence with a $noncritical$  $W_\infty$ string propagating in $AdS_4 \times S^7$.

\bigskip
\centerline{\bf II Euclidean  $AdS_{2n} $ as $SO(2n,1)~\sigma$-model
Instantons }
\bigskip
The Euclideanized $AdS_4$ background will be shown in this section to be an instanton field configuration of 
the conformally invariant $SO(4,1)~\sigma$ model in $R^4$ obeying the
double self-duality condition ( to be described below by eq-(8b)) .    
Conformally invariant Euclidean $SO(5) ~\sigma$-models  in
$R^4$ had a one-to-one correspondence with the conformally invariant $SO(4)$ YM in
$R^4$ whose instanton solutions corresponded geometrically to the stereographic maps of $ S^4$ [1] onto $R^4$. 
It is natural to see whether instanton
configurations of the conformally invariant $SO(4,1)~\sigma$-models in
$R^4$ obeying a double self duality condition exist as well.  
The answer is yes and these correspond to the Euclidean signature $AdS_4$ spaces.

The importance to establish this is because one may generalize these solutions to higher dimensions . For example, 
instantons solutions of the $SO(12, 1)$ conformally invariant  nonlinear sigma-models in 
$ D = 12$ dimensions ( like $R^{12}$ )  
correspond geometrically to 
the Euclidean signature $ AdS_{12}$ spaces.  It is well known that higher dimensional nonlinear sigma models 
are related to branes propagating on curved backgrounds. In particular, $SO(4k)$ Generalized
Yang-Mills (GYM) theories in $R^{4k}$  admit a Moyal deformation quantization whose classical 
$\hbar = 0$ limit furnishes the 
Dolan-Tchakrian actions for $ p$-branes ( $ p+1 =4k$ ) moving in flat backgrounds ( in the conformal gauge) [ 10, 34].   

When the group is compact like $SO(2n)$ , for example, the latter GYM are defined by Lagrangians [1,10] in $R^D$ where $D=2n=4k$ :

$$L=tr~(F^{\alpha_1 \alpha_2....\alpha_{2k}}_{\mu_1,\mu_2....\mu_{2k}}
~\Sigma_{\alpha_1,\alpha_2.....\alpha_{2k}})^2 = (F_{2}\wedge
F_{2}....\wedge F_{2})^2. ~~~2n=4k=D\eqno (6)$$
with all indices antisymmetrized. $\Sigma_{\alpha_1 \alpha_2...}$ is the totally antisymmetrized product ( on the internal indices 
$\alpha$) of the product of $k$ factors of the $2^{2k-1}\times 2^{2k-1}$ matrices $\Sigma_{\alpha_1 \alpha_2}$
corresponding to the chiral representation of $SO(4k)$.

The exhibiting relationship between the instanton field configurations of the $SO(5)~\sigma$ model and the $SO(4)$ YM system 
in $R^4$ 
were given by [1] . 
The order parameter field of the $SO(5)~\sigma$ model is the $SO(5)$ vector $n^i (x)$ obeying the constraint :$n^i (x) n_i (x) =1$. 
The definition of the gauge fields and field strengths for the YM  system in terms of the $n^i(x)$ are :

$$ A^{ij}_\mu =n^i (x) \partial _\mu n^j (x) - n^j (x) \partial _\mu n^i (x) .\eqno (7a)$$

$$F^{ij}_{\mu \nu} (x) =\partial _\mu A^{ij}_\nu - A^{ik}_\mu A^{lj}_\nu \eta_{kl} - \mu \leftrightarrow \nu = 
\partial_{[\mu } n^{[ i} \partial_{ \nu ]} n^{j ]}. \eqno (7b)$$
due to the constraint $n^in_i=1$ the former field stength coincides
with the definition given below by eq-(8c).

An additional constraint allows to reduce $SO(5)\rightarrow SO(4)$ : 

$$D_\mu n^i ( x) = (\delta ^i_j \partial_\mu + A^{ij}_\mu ) n^j (x) =0 . \eqno (7c)$$

After using eqs-(7a,7b,7c) the authors [1] have shown that instanton solutions obey a double self duality 
condition satisfied by the BPST instanton [9] :   :

$$ \epsilon_{\alpha_1 \alpha_2 \beta_1 \beta_2} 
F^{\beta_1 \beta_2 }_{\mu_1 \mu_2} = \epsilon_{\mu_1 \mu_2 \nu_1\nu_2} F^{\nu_1 \nu_2}_{\alpha_1 \alpha_2}\eqno (8a)$$
contrasted with the double self-duality condition exhibited by the $SO(5)~\sigma$ model  in $R^4$ [1] :

$$ \epsilon_{i_1i_2.....i_5} n^{i_5} (x)  
F^{i_3 i_4 }_{\mu_1 \mu_2} = \epsilon_{\mu_1 \mu_2 \nu_1\nu_2} F^{\nu_1 \nu_2}_{i_1 i_2}  \eqno (8b)$$
that stemmed from the $\sigma$ model Lagrangian in $D=4$
$$S=\int d^4x~{1\over 2 (2!)} (F^{ij}_{\mu \nu})^2.~ ~ ~
F^{ij}_{\mu \nu} = \partial_{[\mu}  n^i (x)\partial _{\nu]} n^j (x) 
-\partial_{[\mu}  n^j (x) \partial _{\nu]} n^i (x).\eqno (8c)$$

Similar type of actions are generalized for higher rank objecs :
$$S=\int d^{2n}x~{1\over 2 (n!)} (F^{i_1...i_n}_{\mu_1....\mu_n})^2.~ ~ ~
F^{i_1....i_n}_{\mu_1...\mu_n} = \partial_{[\mu_1}  n^{[i_1} (x).....\partial
_{\mu_n]} n^{i_n]} (x) \eqno (8d)$$

The unit charge instanton solutions of eq-(8b) that minimize the action (8c) and correspond to the $SO(4)$ YM instanton solutions of 
eq-(8a) were found to be precisely the ones corresponding to the stereographic projections ( conformal mappings) from $R^4 \rightarrow S^4$ [1] :
$$n^a (x) = {2 x^a \over 1+x^2}; a=1,2,3,4. ~~~n^5 (x) ={x^2-1 \over
  x^2 +1}.
~~~\sum_{a=1}^{a=4} (n^a)^2 + (n^5)^2 =1. \eqno (9)$$

Things differ now for the (noncompact )$SO(2n,1)~\sigma$-models due to the subtle signature changes. Now we must see whether or not the $SO(2n,1)$ vector $n^i(x)$ with $i=1,2,......2n+1$ satisfy the double self duality condition 
 $$\epsilon_{i_1......i_n i_{n+1}..........i_{2n}.i_{2n+1}} n^{i_{2n+1}} F^{i_{n+1} i_{n+2}.....i_{2n}}_{\mu_{n+1}.........\mu_{2n}} = \epsilon_{\mu_1 \mu_2.....\mu_{n+1}........\mu_{2n} }  F^{\mu_1 \mu_2.......\mu_{n} }_{i_1 i_2......i_n}      \eqno (10)$$

The ``stereographic'' maps from $R^4$ to the Euclidean signature 
$AdS_4$ is defined by :

$$n^a (x) ={2 x^\mu \over 1-x^2};~ a=1,2,3,4.~\mu =0,1,2,3. ~~~n^5 (x) ={1+x^2 \over 1-x^2}. \eqno (11)$$
$$x^2 \equiv (x^0)^2 +(x^1)^2 + (x^2)^2+(x^3)^2.
     \eqno (12)$$
obey the condition : 

$$n^i n_i   \equiv (n^1)^2 +(n^2)^2 + (n^3)^2+(n^4)^2 -(n^5)^2 =-1. \eqno (13)$$
which is just the $SO(4,1)$ invariant norm
of the vector $n^i(x)=(n^1(x),.....n^5(x))$  
and satisfy the double-self duality condition condition (8b) as we
shall see.  

Notice that the solutions ( 11 ) are naturally $singular$  at $ x^2 = 1 $ which correspond to the projective/conformal boundaries of $AdS$ spaces at infinity  
These singularities are naturally due to the noncompact nature of $AdS$ spaces which have $infinite$ volume. 
Hyperbolic manifolds of finite volume are obtained by modding out by a discrete 
subgroup. We shall not be concerned with  these singularity issues in this work since our main concern is to show that the instanton 
solutions given by eqs-(11) are genuine solutions to the double self duality equations. 
Before establishing this fact it is important to study the signature
dependence in the definition of the duality operation.  
For the particlar case  when $2n=4$ ( it can be generalized to any $2n$ ) the double duality star operation
is defined   :

$$^{*} F\equiv { 1 \over (2!)^2 } \epsilon^{\mu_1 \mu_2 \nu_1 \nu_2}\epsilon_{i_1 i_2 i_3 i_4
  i_5} n^{i_5} F^{i_3 i_4}_{\mu_1 \mu_2}. \eqno (14)$$
The $^{**} $ acting on $F$ is :

$$^{**} F=(-1)(-1)^{s_b}(-1)^{s_g}(-1)^f F. \eqno (15)$$
where (i) the first factor of $-1$ stems from the $n^i n_i =-1$
condition. (ii) The second factor stems from the signature of
the base space $R^4 $. (iii) The third factor stems from the group
signature which is $(-1)^1=-1$ for $SO(4,1)$. (iv) The last one,
$(-1)^f$,  are
additional factors resulting from the permutation of the indices in
the $\epsilon_{\mu_1....\mu_{2n}}$ and $\epsilon_{i_1....i_{2n+1}}$
tensors . These are similar to  the 
$(-1)^{r(D-r)}$ factor appearing in the double Hodge operation. Where
$r$ represents the rank of $F$. 
In this case they yield the factor $1$. 

From (15) one immediately concludes that the signature of the base
manifold must be : $(+,+,+,+)$ so that  
 
$$^{**} F=(-1)(-1)^{s_b}(-1)^{s_g} ( -1)^f F=(-1) (+1) (-1)(1) F=F\Rightarrow
(^{*})^2 =1 \Rightarrow ^{*} =\pm 1 . \eqno (16a)$$
and one has a well defined double (anti) self duality condition.  
$$^{*} F=\pm F. \eqno (16b)$$

For higher rank fields, $F^{i_1.....i_n}_{\mu_1.....\mu_n}$ it follows
from (15) that  we must have in addition that $n=even$. Therefore we conclude that
the base manifold must be of Euclidean signature type and have for dimension: $R^{4k}$
and the target group background $SO(4k,1)$,  whose topology is
that of the Euclideanized $AdS_{4k}$ . Thus, the maps are
signature-preserving and the dimensionality of the base manifold must
be a multiple of four,  $D=4k$ . 

It still remains to prove that the $SO(4k,1)~\sigma$-field $n^A (x)$ given by eqs-(11) solve the double self
duality condition (10). In the ordinary $compact$  case, for spheres $S^{2n}$ related to compact $SO(2n+1)$ sigma models,
 the instanton solutions 
saturate the minimum of the action to
ensure topological stability. Also these compact  instanton solutions yield a
finite valued action; i.e the fields fall off sufficiently fast
at infinity so the action does not blow up. 
This will not be the case for the noncompact $SO(2n, 1)$ sigma models discussed here. The $infinite$  volume of 
$AdS_{2n}$  spaces will yield a divergent action despite the fact that the action density is $finite$.
This happens also with the Einstein-Hilbert action for $ AdS$ gravity. It diverges because the $AdS$ volume is infinite. 
But the scalar curvature is everywhere constant and negative despite the fact that the metric $diverges$ at the boundaries.

Physically the $SO(2n, 1)$ instanton solutions realized by the Euclideanized $AdS_{2n}$ spaces represent the $tunnneling$ of a 
closed $p$-brane ( for $ p+1 = 2n$) from $ x_o = - \infty$ to $ x_o = + \infty$. 
The Euclidean signature $AdS_{2n}$ is represented geometrically as the two disconnected branches/sheets  of the 
hyberboloid embedded in $ 2n+1$ dimensions. 
The topology of those two disconnected branches is equivalent to the $2n$-dimensional disk and the metric is 
the Lobachevsky one of constant negative curvature. The instanton solution corresponds physically to a 
closed $ p$-brane ( such $ p+1 = 2n $ ) that starts at $ x_o = - \infty$ and evolves in time shrinking  to  $zero$ size at the point 
$ x_o = - 1 $ ,    
located at the south branch of the hyperboloid , and then $tunnels$ all the way through to  the point $ x_o = + 1$,  
located in the northern branch ,  re-emerging and blowing in size while propagating all the way to $x_o = +\infty$.

To satisfy these requirements it is essential to prove firstly that
the classical action and classical Hamiltonian are positive
definite. A close inspection reveals that one should have an Euclidean
signature base manifold which forces one to choose an Euclidean
signature $AdS_{2n}$ space if the condition (16a) is to be
satisfied. Therefore, instead of $SO(2n-1,2)$ one must have $SO(2n,1)$.  
To illustrate the fact that the classical action is positive definite
and without loss of generality we study the $R^{4,0} \rightarrow
AdS_{4,0}$ case. ( the Euclideanization of $AdS_4$).  We must show
that :

$$(F^{ij}_{\mu\nu})^2 >0.~i,j =1,2,3,4,5.~\mu, \nu =0,1,2,3.  \eqno (17)$$
In $ D=4$ there are six terms of the form : 

$$g^{00}g^{11}[ (F^{ab}_{01})^2 - (F^{5a}_{01})^2 ] +
g^{00}g^{22}[ (F^{ab}_{02})^2 - (F^{5a}_{02})^2 ] +.....$$
$$+ g^{11}g^{22}[ (F^{ab}_{12})^2 - (F^{5a}_{12})^2 ] +.......\eqno (18)$$
Due to the Euclidean nature of the base manifold the $g^{00}g^{11}>0$
so we
must show then that the terms inside the brackets in eq-(18) are all positive
definite. The $a,b$ indices run over : $1,2,3,4$ and the last one is
the $i=5$ component $n^5$ associated with the non-compact
$SO(4,1)$-valued $n^i$ field. 
Setting $n^i = (n^a, n^5 )\equiv (\vec \pi, \sigma)$ the condition
$n^i n_ i = -1$ yields :
$$\sigma^2 = \vec \pi. \vec \pi +1 =\pi^2 +1. \eqno (19)$$
which implies for every single component of $x^\mu; \mu =0,1,2,3$ ( we
are not summing over the $x^\mu$) :

$$(\partial_\mu n^5)^2 < {\pi^2 \over \sigma^2 }(\partial_\mu \vec
\pi).(\partial_\mu \vec \pi);~\mu =0,1,2,3. \eqno (20)$$
Then :
$$(F^{5a}_{01})^2 = {1 \over \sigma ^2}[ (\partial_0  \pi^a )(\vec
\pi . \partial_1 \vec \pi) - (\partial_1  \pi^a )(\vec
\pi . \partial_0 \vec \pi) ]^2 <{\pi^2 \over \sigma^2 }
(\partial _0 \pi^a \partial_1  \pi^b - \partial _1 \pi^a \partial_0
\pi^b) ^2. \eqno (21)$$
where we used  the relation $(\vec A. \vec B)^2 \le A^2 B^2$. The first
term of (18) yields :
$$(F^{ ab}_{01})^2 - (F^{5a}_{01})^2> (1-{\pi^2 \over \sigma^2 })  (\partial _0 \pi^a \partial_1  \pi^b - \partial _1 \pi^a \partial_0
\pi^b) ^2 >0. \eqno (22)$$
because $(1-{\pi^2 \over \sigma^2 })>0$ due to the relation (19). 
The same argument applies for each single component of $(F^{ij}_{\mu
\nu})^2$ and hence the classical action density and Hamiltonian are positive
definite. 

It remains to prove that the solutions (11) obey the double self
duality condition (10). The
latter property holds and can be seen by inspection. The former
property is a straightforward extension of the
results in [1,3] for the compact case. Given
the stereographic  projections defined by eq-(11) : $R^{2n}
\rightarrow H^{2n}$ one can
find a set of independent $2n+1$ frame vectors, $E_p$,  representing the pullback of the
orthogonal frame vectors of the $2n+1$-dim pseudo-Euclidean manifold onto the
Euclidean signature $AdS_{2n}$ space  :

$$E_1=\partial_{x^1} n^i; ~E_2=\partial_{x^2} n^i;..... 
E_{2n}=\partial_{x^{2n}}
n^i.~E_{2n+1}=n^i. ~ i=1,2,.....2n+1\eqno (24)$$
obeying  the orthogonality condition : 

$$g_{\mu\nu}\equiv E^i_\mu E^i_\nu =\partial_\mu n^i \partial_\nu n^i
={4\over (1-x^2)^2} \eta_{\mu\nu}. \eqno (25)$$
which precisely yields the Euclideanized $AdS_{2n}$ metric $g_{\mu\nu}$ definded as
the pullback of the embedding metric, satisfying $R_{\mu\nu}\sim
g_{\mu\nu}\Lambda$ and $R\sim \Lambda$. The latter curvature relations
are the hallmark of Anti de Sitter space ( negative scalar curvature). 
The positive definite measure spanned by the collection of $2n+1$ frame vectors 
defines the natural volume element of the Euclideanized $AdS_{2n}$ :

$$\sqrt {det ~h} = det~(E_1,E_2,......,E_{2n}, E_{2n+1} ). \eqno (26)$$

It was shown by [1,3] (for the compact case the orthogonality
condition is naturally modified to yield the metric of the $S^{2n}$) that the orthogonality conditions eqs-(25) are
the $necessary$ and $sufficient$ conditions to show that the
particular field configurations $n^i(x)$ given by
(11), after extending the four dimensional result to any $D=2n$,
satisfy the double self duality conditions eq-(10). 
The orthogonality relation implies that the angles among any two vectors is
preserved after the mapping and also that all the vectors , after the mapping, are of equal length  
( this is not to say that their 
length equals the same as before the mapping).
This is a signal of a conformal mapping. The ``stereographic'' maps are
a particular class of conformal maps. The action of the conformal
group in $R^{2n}$ will furnish the remaining conformal maps [1,3]. 
Hence, we see here once again the interrelation between self-duality
and conformal invariance. In two dimensions self-duality implies
conformal invariance ( holomorphicity). In higher dimensions matters
are more restricted, they are not equivalent.

Therefore, setting aside the subtleties in the quantization program
due to the noncompact nature of $SO(2n,1)$,  we have shown that these
$SO(2n,1)~\sigma$-model instanton field configurations obtained by
means of the ``stereographic'' ( conformal) maps of $R^{2n}
\rightarrow H^{2n}$, correspond precisely to the coordinates,
$y^1,y^2,....y^4,y^5$ of the  $5$-dim pseudo-Euclidean manifold onto
which the Euclidean signature $AdS_4$ space is embedded. 
To illustrate this lets focus on Euclidean signature $AdS_4$ defined as  the geometrical locus :

$$y^2 \equiv (y^1)^2 +(y^2)^2 + (y^3)^2+(y^4)^2 -(y^5)^2 =
-\rho^2=-1. \eqno (27)$$
where we set the $AdS$ scale $\rho=1$. 
The $AdS_4$ coordinates : $z^0,z^1,z^2,z^3$ are related to the $y^A$
using the ``stereographic'' projection ( see the lectures by Petersen
[6]) from the ``south'' pole of the Euclidean signature $AdS_4$ 
to the equator $R^{2n}$ passing through the orgin, dividing
 $R^{2n+1}$ 
into a north/south region, and finally intersecting the upper branch of the hyperboloid at point $ P $  :

$$y^a  (z) =\rho{2 z^\mu \over 1-z^2};~\mu=0,1,2,3. ~~~a=1,2,3,4. 
~~~y^5 (z) =\rho {1+z^2 \over 1-z^2}. \eqno (28)$$
it is straighforward to verify that $y^2=-\rho^2=-1$. 
The Euclidean signature $AdS_4$ metric is equivalent to the Lobachevski metric of a Disk; it is 
conformally flat :

$$g_{\mu \nu} = {4 \over (1-z^2)^2} \eta_{\mu \nu}¸~~~ z^2 =(z^0)^2
+....+(z^3)^2. \eqno (29)$$ 
This conformally flat metric can be generalized to any dimensions and have 
constant negative curvature given by $ R = - { d (d-1)/a^2 } $ where the Disk's  radius is $ 2a$.   
There is an exact  match with the instanton field configurations $n^i
(x)$ (11). The metric (29) also matches (25) 
after the correspondence : $n^a (x)\leftrightarrow y^\mu (z)$ ; 
$n^5 (x) \leftrightarrow y^5 (z)$  and $ x^\mu \leftrightarrow z^\mu $ 
is made because the $z^\mu$ coordinates used by Petersen [6] 
can be identified precisely with the $x^\mu$ coordinates of $R^4$.  

Therefore,  concluding, the Euclidean $AdS_4$ space can indeed be seen as 
the target space of a nonlinear $SO(2n,1)$ sigma model defined in $ R^{2n}$ obeying the double self duality conditions ( 8 ).   
Since these instanton  solutions  are determined $modulo$ conformal transformations, for example, 
one can conformally map the $4$-dimensional disk into the upper half plane $R^{4 }$  and in these  
Poincare coordinates the $AdS$ metric reads : 

$$ ds^2 = { a^2 \over (x^4)^2 } [ ( dx^1)^2 + (dx^2)^2 + ( dx^3)^2 + (dx^4)^2 ].  ~~~x^4 > 0.                    $$ 
In $ d = 2 $ it coincides with the familiar metric in the upper half of the complex plane. 
This four dimensional result can be generalized to higher dimensions in a straightforward fashion.

These conformally invariant $SO(2n,1)$ sigma model actions in $R^{2n}$
are just the sigma model generalizations of Yang-Mills type actions. 
The topological Chern-Simons actions are related to the MacDowell-Mansouri-Chamsedinne-West action for
ordinary gravity based on gauging the conformal group $SO(3,2)$ in
$D=3$ Minkowski space which is also the same as the isometry group of $AdS_4$ . 
The ordinary Lorentz spin connection and the tetrad in four dimensions
are just pieces of the $SO(3,2)$ gauge connection. Upon setting the
torsion to zero one recovers the $4D$ Einstein-Hilbert action with a
cosmological constant and the Gauss-Bonnet topological term. 
The conformally invariant sigma models actions [1] are the ones required in building
conformally invariant bosonic $p=2n$-brane actions in flat or curved target backgrounds. It would be
interesting to see what sort of actions can be derived following the
analog of a MacDowell-Mansouri procedure to obtain Einstein gravity from a gauge theory. 
And this is precisely what we shall do in section {\bf 4}.

\bigskip

\centerline {\bf III } 
\bigskip 
\centerline{\bf 3.1 Self-Dual $p$-branes, Chern-Simons $p'$-Branes }
\bigskip

In the first part of this section we will study the relationship between Self-Dual $p$-branes and Chern-Simons $p'$-Branes, 
for $ p'+1 = p = D-1$  
provided the target embedding $D$-dim spacetime is ( pseudo) Euclidean .  
In the second part we will review our work [33-35] and show  how ( Chern-Simons ) brane actions
emerge from the large $N$ limit of quenched, reduced ( Generalized ) Yang Mills theories in diverse dimensions.  
Also we will show how the large $N$ limit of Matrix Chern-Simons theories are also given by Chern-Simons branes. 

Having shown in {\bf 2 } how Euclidean signature $AdS_{2n}$ spaces are instanton solutions of the 
conformally invariant $SO(2n,1)$ sigma models in $ D = 2n$ dimensions 
we turn our attention to the construction of self dual branes moving in $ 2n$ dimensional backgrounds . 
The latter self dual  branes obey a similar $double$ self duality condition 
as the former instantons . 
It was pointed out in the introduction that Zaikov had shown that self-dual $p$-branes, 
when $p+1=D=2n$ and the embedding $2n$-dimensional manifold is  ( pseudo ) Euclidean  [11], 
are related to Chern-Simons $p'$-branes, for $p'+1=p=2n-1 $  
which implied that Chern-Simons  branes must have codimension $two$ : 
$ p'+2 = 2n $ . 

In essence, this correspondence says that the world volume of a boundary matches 
the boundary of a world volume. This is $not$  the same as saying that the Self Dual $p$-brane 
is itself the world volume of a Chern-Simons $p'$-brane. 
For this to happen, it will require a theory 
with $two$ times : One time representing the world volume dynamics of the Chern-Simons $p'$-brane 
and another time representing the world volume dynamics of the $p$-brane.  
Two-times physics have been amply studied by Bars [26].   
For this reason that Chern-Simons branes have codimension $ two$ they 
are the $high$ dimensional analogs of $knots$  [47] . 
In particular we must distinguish the following cases in this work :

$\bullet $ The Euclidean signature $AdS_{2n}$ spaces were interpreted as the tunneling of a 
closed $p$ brane  ( $p = 2n-1$  ) from $x_o = -\infty$ 
( lower sheet of the hyperboloid ) 
to $x_o = + \infty$ ( the upper sheet of the hyperboloid ) 
after shrinking to $zero$  size at point $ x_o = - 1$ and re-emerging at point $x_o = + 1$. 
The relevant self-dual branes we will study in this section 
are related to those Chern-Simons branes living on the boundaries of a $ D$-dimensional region as 
we shall see next. 

$\bullet$ $AdS_{2n}$ spaces have the topology $ S^1 ( time ) \times R^{ 2n -1} ( space )$ and 
the latter space has the same topology as the 
lateral component of the $boundary$ of a space of topology $ D^2 \times R^{2n-1}$ where 
$D^2 $ is the two-dim disk. The $covering$  
space of $AdS_{2n}$ has the topology of  $ R^1 \times R^{ 2n -1} = R^{2n }$ 
which in turn can be thought of as the 
world volume of an $open$ brane of topology $ R^{2n-1}$ such $ p' = 2n-1$. 
This open $2n-1$-brane can be identified as the 
Chern-Simons brane of codimension $two$  :  $p'+2 = (2n-1) + 2 = 2n+1 $ ; 
i.e it is the embedding of an $2n-1$ dimensional open manifold 
of topology $R^{2n-1}$ onto the 
pseudo Euclidean manifold of dimension $ D = 2n+1 $. 
This is the orginal pseudo-Euclidean manifold where the $AdS_{2n} $ and the $2n$-dim 
de Sitter spaces were naturally embedded.    
Hence, from the $covering$ space perspective of $AdS_{2n}$ , 
the $open$  Chern-Simos $p'$-brane ( for $p' = 2n-1$) lives on the 
$covering~ space$  of the lateral $boundary$  component of $ D^2 \times R^{2n-1}$; i.e.  
It lives on the covering space of $ AdS_{2n}$.

However, this does $not$ mean that $AdS_{2n}$ spaces will generate the $ 2n+1$-dimensional  
( pseudo) Euclidean spaces $R^{2n+1}$ onto which they are being embedded. 
Since $AdS_{2n} $ spaces are $curved$ , their would-be world-volumes of topology $ AdS_{2n}  \times R^1 $ 
would be curved also , hence this disqualifies  the flat $R^{2n+1}$  as its world volume.  
For this reason $AdS_{2n} $ spaces are $not$  Self Dual $ 2n$-branes living in flat $2n+1$ dimensions.   
Notice also, that based on the results of the previous section, 
the $AdS_{2n}$ spaces $cannot$ be seen as  
instanton solutions of the $ SO(2n-1, 2 )$ non-linear $\sigma$ model, because 
they don't have the appropriate signature, like the Euclideanized  $AdS_{2n}$ spaces did. 
The correct  signature was an essential 
condition to find real valued solutions to the double self duality equations ( 8 ) . 
Hence, it is very important to keep in mind the difference beween the Euclidean $AdS_{2n} $ instantons solutions 
of the conformally invariant nonlinear $SO(2n , 1)$ sigma model in $ D = 2n$ dimensions  
and the actual $ AdS_{2n}$ spaces.

$\bullet $ Another very relevant case is  the odd dimensional $AdS_{2n-1} $ space. It has the topology  
$ S^1 ( time ) \times R^{ 2n -2 } ( space )$ and the latter space has the same topology as the 
lateral component of the $boundary$ of a space of topology $ D^2 \times R^{2n-2}$ where 
$D^2 $ is the two-dim disk. The $covering$  
space of $AdS_{2n-1}$ has the topology of  $ R^1 \times R^{ 2n -2} = R^{2n-1 }$ which in turn can be thought of as the 
world volume of an $open$ brane of topology $ R^{2n-2}$ such $ p' = 2n-2$. 
This open $2n-2$-brane can then be identified as the 
Chern-Simons brane of codimension $two$ :  $p'+2 = (2n-2) + 2 = 2n  $ ; 
i.e it is the embedding of an $2n-2$ dimensional open manifold of topology $R^{2n-2}$ onto the (pseudo)  Euclidean manifold of 
dimension $ D= 2n $.

In particular, to see now the specifics of Zaikov's 
self dual $p$- branes and Chern-Simons $p'$ branes , when $ p'+ 1 = p = 2n-1$ ; i.e. when 
the world volume of a boundary matches the boundary of a world  volume, 
we shall write down the 
Dolan-Tchrakian action for a spacetime filling bosonic
$p$-brane  ( with $p+1=2n = D $ ) embedded in a flat target spacetime of
dimensionality $D=p+1=2n$.  The conformally invariant Dolan-Tchrakian action for $p$ branes such that $ p+1 = 2n$ are  
in general valid for both flat or curved target spacetime backgrounds [2].  The action in $flat$ backgrounds can be written 
in this form  : 

$$S = T \int d^{2n}\sigma~ \sqrt h ~ h^{a^1 b^1}....h^{a^n b^n} ~
\partial_{[a^1} X^{\mu_1}\partial_{a^2} X^{\mu_2}...\partial_{a_n] } X^{\mu_n}  
\partial_{[b^1} X_{\mu_1}\partial_{b^2} X_{\mu_2}...\partial_{b_n] } X_{\mu_n} . \eqno (30a)$$
where the antisymmetrization of the world-volume indices is explicitly shown. 
$T$ is the brane tension of units of $(mass)^{2n}$ and the target spacetime dimension is $ D \ge 2n$. 

Such action in $flat$ backgrounds can also be rewritten explicitly 
in the Yang-Mills like $F\wedge ^{*} F = F_{\mu\nu}F^{\mu\nu} $  form  where one just needs to enlarge 
the indices of the field-strengths to accomodate anti-symmetric tensors of even rank $ n$. It was precisely 
derived based on the conformally invariant $SO(2n+1)~\sigma$-model actions given by eqs-(8) : 

$$S = T \int 
[E^{p_1}\wedge E^{p_2}\wedge...\wedge E^{p_n}]~\wedge ~ ^{*} 
[ E^{q_1}\wedge E^{q_2}\wedge....\wedge E^{q_n} ]\eta_{ p_1 q_1}\eta_{p_2q_2}..... \eta_{p_n  q_n} \eqno (30b)$$

The star operation in (30) is the standard Hodge duality one defined w.r.t the $p+1$-dim
worldvolume  metric  of the $p$-brane : $h_{ab}$. 
For branes that saturate the embedding dimensions ( $ p+1 = 2n + D $ ) 
The world-volume one-forms ,
$E^p;~p=1,2,....2n$ are obtained as the pullbacks of the $D=2n=p+1$ one
forms, $e^A~;A=1,2,....2n$ 
associated with the $D=2n$ dimensional target space onto which we embed the
world-volume of the $p$-brane.  $\eta_{pq}$ is  the tangent space ( to the world-volume) Minkwoski metric obeying 
$ E^{p}_a E^{q}_b \eta_{pq} = h_{ab} $ .

A self-dual $p$-brane ( self dual w.r.t the Hodge star
operation associated with the worldvolume metric ) obeys :

$$^{*}
[E^{q_1}\wedge E^{q_2}\wedge....\wedge E^{q_n}]= 
 [E^{q_1}\wedge E^{q_2}\wedge...\wedge E^{q_n}]. \eqno (31)$$
and will automatically satisfy the equations of
motion. This is the $p$-brane generalization of the relation
$d^{*}F=^{*}J=0$ in the absence of sources in the YM equations. When
$^{*}F=F$ the latter equations become the Bianchi identities $dF=0$. 

Hence, if the self duality condition eq-(31) is obeyed , 
the action ( 31 ) becomes in this case : 

$$ S = T \int E^{p_1}\wedge E^{p_2}\wedge...\wedge E^{p_n} \wedge E^{q_1}\wedge E^{q_2}\wedge...\wedge E^{q_n}  . $$ 
and the action (30) for a self-dual $p$-brane becomes $precisely$ the integral
over the ordinary Jacobian ( a $p+1=2n$ volume form) associated with the change
of variables $\sigma \rightarrow X$ and hence it is topological.  
It has no transverse degrees of freedom since 
$ p+1 = 2n = D$.  If the world volume has a natural boundary,
an integration by parts ( assuming one has a valid global coordinate patch )  
will yield the Chern-Simons $p'=p-1$-brane as indicated in eq-(4,5). 
In this way we have seen then how a self-dual $p$-brane is indeed related to the Chern-Simons $p'=p-1$-brane ( at
least on shell) and it has codimension $two$ : $ p'+2 = p+1 = 2n = D$. 

In the particular case of a $3$-brane , the Dolan-Tchrakian action in $flat$ backgrounds is  

$$ Tr ~ F \wedge ^{*} F \leftrightarrow 
\int d^4 \sigma ~ \{ X^\mu , X^\nu \}_{PB} \{ X_\mu, X_\nu \}_{PB}   $$
And for a self-dual $3$-brane ,  
one has the brane analog of the Self Dual YM relation : 
$$ ^* F = F \Rightarrow  F \wedge ^{*} F = F\wedge F $$ 
and the Dolan-Tchrakian action becomes :    
$$ Tr~ F \wedge F \leftrightarrow 
\int d^4 \sigma ~  \epsilon_{\mu\nu\rho\tau }          \{ X^\mu , X^\nu \}_{PB} \{ X^\rho, X^\tau \}_{PB}  $$     
and it is purely  topological , a pure volume term.  An integration by parts ,
assuming one has a well defined and unique global coordinate chart that covers the whole four-dim domain,  
it correspond to the $3$-dim world volume associated with a 
Chern-Simons membrane living on the $3$-dimensional boundary of the four-dimensional domain.

Before finishing this section we will concentrate on what are the requirements 
for the self-duality conditions from the world-volume perspective to agree with 
the self duality conditions from the spacetime point of view and with the   
double self-duality conditions formulated by Zaikov when the embedding manifold is
Euclidean. For example,  when $p+1=4$, the double self duality equations of Zaikov can be written [11] : 

$${\cal F}^{JK}_{ab} = 
\partial_{\sigma^a} X^J \partial_{\sigma^b} X^K-\partial_{\sigma^b} X^J \partial_{\sigma^a} X^K
={1\over 4}\epsilon _{abcd}\epsilon^{JKLM} {\cal F} ^{LM}_{cd}. \eqno
(32a)$$
where $X^I (\sigma^a)$ are the embedding coordinates of the $p$-brane
into the target space, 
$X^I;~I=1,2,..4$. It was obtained from the standard Dirac-Nambu-Goto actions. 
The Dolan-Tchrakian action (30) is equivalent at the classical level to
the Dirac-Nambu-Goto actions upon elimination of the auxiliary
world-volume metric, $h_{ab}$. Whether or not this equivalence occurs Quantum
Mechanically is another issue.  Because the world-volume of a spacetime filling branes can be identified with ( a subset of ) 
the target background, one has at the classical level , that Zaikov's
double  self-duality conditions (32) on $Euclidean$  embeding spacetime backgrounds must have a one-to-one correspondence 
to the self-duality conditions (31) associated with $Euclidean$  world-volumes of the 
self-dual branes furnished by Dolan-Tchrakian actions (30). We will see this next. 

Before doing this one should notice that in eq-(32a) one could replace $upper$  
spacetime tensor indices for $lower$  spacetime tensor indices  
since the embedding manifold is supposed to be 
Euclidean; i.e eq-(32a) is indeed invariant under such replacement : 
$X^J \rightarrow X_J$ since : $X^J = \eta^{JM}X_M = X_J$
so this replacement does $not$  affect the outcome of the form of eqs-(32a) . 
Since Zakov's equations were written in 
Euclidean backgrounds it is more approriate to write the double self duality equations in the more general form 
that are also valid in pseudo-Euclidean spacetimes and that bear a closer 
resemblence to the double self duality equations of the 
non-linear sigma models (8). 
Thus the  double  self duality equations of Zaikov (32a) in ( pseudo ) Euclidean manifolds should have the 
precise index-position  : 

$$  {\cal F}^{JK}_{ab}  = {1\over 4} \epsilon_{abcd} \epsilon^{JKLM}  
[ \partial^{c} X_L \partial^{d} X_M - \partial^{d} X_L \partial^{c} X_M ]    
= ^* [ {\cal F}_{LM}^{cd} ].    \eqno (32b)$$

To arrive at the ordinary self-duality equations ( in the Yang-Mills like form ) 
from the double-self-duality ones we must multiply  both sides of eq-(32b) by the inverse of the  
nondegenerate symplectic two-form $ \omega^{ab}$ and sum  over all the 
world-volume indices .  
Upon doing so one can rewrite (32b) in the Yang-Mills like form in terms of the Poisson brackets 
w.r.t. the world  volume coordinates defined as : 

$$ \{ A, B \} \equiv \sum_a ~\omega^{ab} \partial_a A (\sigma ) \partial_b B (\sigma)  \Rightarrow  $$
$$  \{ X^J, X^K \}_{PB}  = \epsilon^{JKLM} \{ X_L, X_M \}_{PB}   \leftrightarrow F^{JK} = \epsilon^{JKLM} F_{LM}.     \eqno (32c) $$ 
This was due to the fact that one can choose a basis 
such that the four-world-volume form $\Omega_{4} $ and the symplectic two-form obey : 
$$ \Omega_{4} = \omega \wedge \omega \Rightarrow \epsilon^{abcd } = \omega^{ab} \wedge \omega^{cd }. ~~~ 
\omega^{ab} \omega_{ab} =  Trace~ I_{ 4 \times 4 }  = 4 .  \eqno ( 32 d ) $$

The reader may ask : why the use of a symplectic two-form and Poisson brackets ? 
In section {\bf 3.2} we will justify the use of 
Poisson brackets directly from the large $N$ limit of $SU(N)$ YM theories.  
Following similar arguments as those given in section {\bf 2 } pertaining the signature 
subtleties of these more general double  self duality equations (32), 
for $even$ rank forms,  in $D = 2n = 4k $ dimensions ,  
it follows that the signature of the worldvolume of the $p$-brane ( $ p+1 = 2n$)   
has to $match$  naturally the target spacetime signatures if we wish to have $real$-valued solutions for $ X^\mu (\sigma)$.  

Furthermore, one can notice also that the (double)  self-duality conditions of $p$-branes resemble those of 
ordinary $SO(4)$ Yang-Mills where the ensuing relation between the two equations is 
implemented after the correspondence  :

$$A^{\alpha \beta}_\mu \Sigma_{\alpha \beta} \leftrightarrow X^J (\sigma^a). ~~~F^{\alpha \beta}_{\mu \nu} \Sigma_{\alpha \beta}
\leftrightarrow 
{\cal F}^{JK}_{ab}= \partial_{\sigma^a} X^J \partial_{\sigma^b}
X^K-\partial_{\sigma^b} X^J \partial_{\sigma^a} X^K. \eqno (33)$$
where $\Sigma_{\alpha \beta}$ are the $2\times 2$ matrices ( Pauli)
corresponding to the chiral representation of $SO(4)\sim SU(2)\times
SU(2)$. The $\mu,\nu..$ indices in the l.h.s of (33) run over the
four dimensional base manifold $R^4$ where the $SO(4)$ YM fields
live. The $I,J,K...$ in the r.h.s of (33) run over the
four-dimensional embedding target space . The $\sigma^a$ indices run over the
four-dimensional world volume of the $p=3$-brane. 
In the next sections  we will show rigorously the validity of such correspondence ( 33 ) 
via the Moyal deformation quantization and why the use of Poisson brackets.

Therefore, the self-duality conditions of the  ( spacetime filling ) Dolan-Tchrakian
$p$-brane actions  in ( pseudo ) Euclidean embedding manifolds ( for the case
$p+1=4$)  (31) have a one-to-one correspondence with the 
double  self-duality conditions (32b) for $p$-branes and ,
 also, with the double
self-duality conditions of the $SO(4)$ YM instanton (8a), if, and only
if, the gauge fields/target spacetime coordinates  correspondence (33)
is used. This gauge-fields/target spacetime coordinates $correspondence$ is precisely the one we will use in 
the next sections to show 
how the large $N$ limit of quenched reduced QCD admit hadronic bag excitations. 
Furthermore, these results can be generalized to higher 
dimensional cases when : $ p+1 = 2n = 4k $.

Before ending this subsection we will discuss the Chern-Simons $10$-brane connected to $AdS_{11}$ spaces. 
As we said earlier, because $ AdS_{11}$ is  $not$ flat it would be inconsistent to claim that it also qualifies as 
a self dual $11$-brane moving in a $flat$  pseudo- Euclidean $12$-dim background.     
The topological Chern-Simons action for the relevant  open $p'=10$-brane is  
defined over the  $covering$ space of the  lateral $d=11$-dimensional boundary component  of a domain of topology 
$D^2 \times R^{10}$ .  This Chern-Simons open $10$-brane has the topology of  $ R^{10}$.  
Its $curved$ world volume can be identified  with the $covering$ space of $AdS_{11}$, since   
such covering has the topology of $ R^1 \times R^{10}$ 
and a non-Euclidean metric of constant negative scalar curvature, conformal to that of a half cylinder, an Einstein universe.  
Since the Chern-Simons $p=10$-brane theory is topological,
it has no local $ 12$-dimensional bulk degrees of freedom, only global, that are naturally
confined to the $11$-dimensional $boundary$. 
In this sense the theory is holographic [27,28] and it is no wonder why  the coverings of 
$AdS$ spaces are of great importance to implement the holographic principle [6].

A final word of  caution : The boundary dynamics of this Chern-Simons $p'= 10$ brane example must 
$not$ be confused with the dynamics associated with the other 
components of the boundary of the domain of topology $ D^2 \times R^{10}$; 
namely the $lids$   at $infinity$ whose topology is  $ D^2 \times S^9$.  This is consistent with the fact that 
the projective boundary of $AdS_{11}$  ( at infinity )  has the topology of $ S^1 \times S^9$. 
Singleton/Doubleton Conformal Field Theories in 
$ S^1 \times S^{p} $  have been analyzed in [ 51, 52  ] 
in connection to the massless spectrum of supersymmetric  $p$-branes propagating on 
$AdS_{p+2} \times S^{ D-p-2 }$  backgrounds.   
These singleton/doubleton  field theories 
are tightly coonnected with $p$-branes living at the " end " of the universe ; 
i.e their world volumes are $ p+1$ dimensional, 
corresponding to the projective boundaries ( at infinity )  of $ AdS_{p+2} $ spaces,   of topology $ S^1 \times S^p$ , 
and whose $covering$  space has the topology of $R^1 \times S^p$ , as it should, to qualify for a world volume of a 
closed $p$-brane of topology $S^p$.

Since we have $d=10, 11,12$ for characteristic dimensions, this particular example may be relevant to understand  
some of the intricate relations of $M,F,S$ Theory  [25,26] and to shed some light into the geometrical, 
holographic and topological underpinnings behind the $AdS/CFT$  duality conjecture.  
These example can be generalized to the other higher/lower  dimensions and to the 
$ AdS_{12}$ case where the relevant dimensions are then $11, 12, 13 $.  
Higher dimensional topological actions have been proposed by Chapline
to describe unique theories of gravity and matter [24].

\bigskip

\centerline { \bf 3.2   Hadronic Bags and Chern-Simons Branes from Quenched } 
\centerline{\bf Large N QCD  in Flat Backgrounds } 
\bigskip

This section deals with the importance that Deformation quantization has in the construction of ( Chern-Simons ) branes from 
the large $N$ limit of $SU(N)$ YM in flat backgrounds. 
$SU(N)$ reduced, quenched gauge theories have been  recently  shown  by us to be related to Hadronic Bags 
and Chern Simons Membranes in the large $N$ limit [33-35] . 
This is reminiscent of the chiral model approaches to
Self Dual Gravity based on Self Dual Yang Mills [31] theories. A Moyal
deformation quantization of the Nahm equations associated with a $SU(2)$ YM [32] theory yields the $classical$~$N\rightarrow
\infty$ limit of the $SU(N)$ YM  Nahm equations  $directly$,  without ever using the 
$\infty \times \infty$ matrices of the large $N$ matrix models. By simply taking the
classical $\hbar =0$ limit of the
Moyal brackets, the ordinary Poisson bracket algebra associated with
area-preserving diffs algebra $SU(\infty)$   is  automatically recovered.   
This supports furthermore the fact that the underlying geometry may be
noncommutative [18,32] and that $p$-branes are essentially gauge theories of
area/volume...preserving diffs [14,20,21].     

This Moyal deformation approach also furnishes dynamical membranes as well [35]  when the  
spatial quenching approximation to a line ( one dimension ) ,  instead of a point ,   is used. 
Basically, a  Moyal quantization takes
the operator ${\hat A}_\mu (x^\mu)$ into $A_\mu(x^\mu;q,p)$ and
commutators into Moyal brackets. A dimensional reduction to one
temporal dimension brings $A_\mu (t,q,p)$, which precisely corresponds
to the membrane coordinates $X_\mu (t,\sigma^1,\sigma^2)$ after
identifying $\sigma^a$ with $q,p$. The $\hbar =0$ limit turns the Moyal
bracket into a Poisson one. It is in this
fashion how the large $N~ SU(N)$ matrix model bears a direct relation to the 
physics of membranes. The Moyal quantization explains this in a
straightforward fashion without having to use $\infty \times \infty$
matrices ! Membrane actions have also been given by Dolan-Tchrakian in terms of 
nonlinear sigma models. The crucial difference is that these sigma models are no 
longer conformally invariant since the world volume dimension is $odd$.  
They have the similar form as the Skyrme action. 
The quadratic terms are the kinetic ones and the quartic derivative terms play the role of a potential term. 
The relevance of Skyrme models in the construction of the efective action for $SU(2)$ QCD has been pointed out recently by [ 69].

We will review how a $4D$ Yang-Mills theory  reduced and quenched, and supplemented by a topological $theta$ term can be related 
through a Weyl-Wigner Groenowold Moyal ( WWGM) quantization procedure to an $open$  domain of the 
$3$-dim disk $D^3$; i.e. an open $3$-brane model [ 33 ]. 
The bulk $ D^3 \times R^1 $ is the interior of a hadronic bag and the 
(lateral ) boundary is the Chern-Simons world volume $ S^2 \times R^1  $ 
of a membrane of topology $ S^2 $ ( a codimension two object ).  
Hence, we have an example where the world volume of a boundary   $ S^2 \times R^1 $  
$is$ the lateral  boundary of the  world volume of an open $3$-brane of topology $ D^3 $  : 
 $ \partial ( D^3 \times R^1 )  = S^2 \times R^1 $ ( setting asside the points at infinity). 
The boundary dynamics $is ~not~  trivial$ despite the fact that there are no transverse bulk 
dynamics associated with the interior of the bag.  This is due to the fact that the $3$-brane is spacetime filling : $ 3+1 = 4$ 
and therefore has no transverse physical degrees of freedom. 
The reduced quenched action in $ D = 4 $ is : 

$$ S = - { 1\over 4} ( { 2\pi\over a } )^4 { N \over g^2_{YM}} Tr~ F_{\mu\nu} F^{\mu\nu } . ~~~~ F_{\mu\nu} = [ i D_\mu, iD_\nu ]. \eqno ( 34 ) $$

The reduced-quenched action is defined  at a " point "  $x_o$ . 
The quenched approximation is based in  replacing the 
field strengths by their commutator dropping the ordinary derivative terms. 
For simplicity we have omitted the matrix $SU(N)$ indices in ( 34 ).  The $theta$ term is : 

$$   S_\theta  = - { \theta N g^2_{YM}  \over 16 \pi^2   } ( { 2\pi\over a } )^4  \epsilon^{\mu\nu\rho \sigma} Tr~ F_{\mu\nu} F_{\rho \sigma  }. 
\eqno (35 ) $$

The WWGM quantization establishes a one-to-one correspondence between a linear operator $D_\mu = \partial_\mu + A_\mu $ acting on the Hilbert space ${\cal H} $ of 
square integrable functions in $R^D$ and a smooth function ${\cal A}_\mu (x, y)$ which is the Fourier transform of $ {\cal A}_\mu (q, p)$. 
The latter quantity is obtained by evaluating the 
trace of the $D_\mu = \partial_\mu + A_\mu $ operator 
summing over the diagonal elements with respect to an orthonormal basis in the Hilbert space.  
Under the WWGM correspondence , in the quenched approximation, the matrix product $ A_\mu . A_\nu $ is mapped into the 
$noncommutative $ Moyal star product of their symbols 
${\cal A}_\mu * {\cal A}_\nu $ 
and the commutators are mapped into their Moyal brackets : 

$$ { 1 \over \hbar } [ A_\mu, A_\nu ] \Rightarrow \{ {\cal A}_\mu, {\cal A}_\nu \} . \eqno ( 36 ) $$ 

Replacing the Trace operation with an integration w.r.t the internal phase space variables , $\sigma \equiv q^i, p^i$ : 

$$  { (2\pi )^4 \over N^3 } \rightarrow \int d^4\sigma . \eqno ( 37 ) $$

The WWGM deformation quantization of the quenched-reduced orginal actions is : 

$$ S^* = - { 1\over 4} ( { 2\pi\over a } )^4 { N \over g^2_{YM}} \int d^{ 4} \sigma ~ {\cal F} _{\mu\nu} (\sigma )   *  {\cal F} ^{\mu\nu } (\sigma ) . ~~~~ 
{\cal F} _{\mu\nu} = \{ i {\cal A}_\mu,  i {\cal A}_\nu \} . \eqno ( 38 ) $$

And the corresponding WWGM deformation of the $theta$ term : 
$$   S^*_\theta  = - { \theta N g^2_{YM}  \over 16 \pi^2   } ( { 2\pi\over a } )^4 \epsilon^{\mu\nu\rho \sigma} \int d^{4}
\sigma ~ {\cal F}_{\mu\nu} (\sigma ) * {\cal F}_{\rho \sigma  } (\sigma ) . \eqno ( 39) $$

By performing the following gauge fields/coordinate correspondence : 

$$ {\cal A}_\mu (\sigma ) \rightarrow ( {2\pi \over N}  )^{1/4} X_\mu (\sigma ) .~~~ {\cal F}_{\mu\nu} (\sigma) \rightarrow 
( {2\pi \over N} )^{1/2} \{ X_\mu (\sigma) , X_\nu (\sigma ) \} . \eqno ( 40  ) $$ 

And, finally, by setting the Moyal deformation parameter $ " \hbar " = 2\pi / N$ 
of the WWGM deformed action (38), to  zero ; i.e taking the classical $\hbar = 0$ limit , 
which is tantamount to taking the $ N = \infty$ limit, 
one can see that the action (38) will become the Dolan-Tchrakian action ( 30 )   
for a  $3$-brane, in the conformal gauge : 
$ h_{ab} = e^{2 \chi (\sigma)} \eta_{ab} $,  moving in a $flat$  $ D =4$-dim background   : 

$$  S = - { 1\over 4 g^2_{YM} } ( {2\pi \over a}   )^4  \int d^4\sigma ~   
\{X^\mu, X^\nu \}_{PB} \{X^\rho , X^\tau \}_{PB} ~ \eta_{\mu\rho} \eta_{\nu \tau }    
   \eqno ( 41 ) $$ 
due to the fact that the Moyal brackets collape to the ordinary Poisson brackets in the $\hbar = 2\pi/N = 0$ limit ( large $N$ limit ) .  

Whereas the action corresponding to the $theta$ term ( 39) 
will become in the $ N = \infty$ limit, the Chern-Simons Zaikov action for a closed membrane embedded in a 
four-dimensional ( pseudo ) Euclidean background 
and whose $3$-dim worldvolume is the boundary of the four-dim hadronic bag .  
The Chern-Simons membrane has $nontrivial$ boundary dynamics compared with the 
trivial bulk dynamics of the spacetime filling $3$-brane. 

Upon identifying the inverse lattice spacing of the large $N$ quenched, reduced {\bf QCD} with the scale of 
$ ( 2 \pi / a ) = \Lambda_{QCD} = 200~ Mev $ 
we were able to obtain the value of the dynamically generated bag constant, $\mu$,  of mass dimensions, related to the bag Tension , 
as :

$$ T =  \mu^4  = { 1\over 4 \pi } ( 200~ Mev)^4 { 1 \over (g^2_{YM}/4\pi)} . \eqno ( 42) $$ 

Taking for the conventionally assumed value of $ g^2_{YM}/4 \pi \sim 0.18 $ , 
the actual value of the hadronic bag constant $\mu$ turns out to be also close to the value 
$\Lambda_{QCD} = 200~ Mev \sim ( 2\pi / a )  $ 
which falls within the range of the phenomenologically accepted values $ \mu \sim 110 ~ Mev $ if we take into account the 
uncertainty on the value of the $ \Lambda_{QCD} $ which lies in the range of $ 120~ Mev$ and $ 350~ Mev$. 

To prepare the territory for the next section, when we introduce gravity into the picture we will set the inverse lattice spacing $ a$ to be $ a^4 = N L^4_{Planck} $. 
In the next section we will $justify$  the validity of this relation. Upon inserting this value of $ a^4$ into eq-( 42 ) and 
recalling that $\mu \sim a^{-1} $ : 

$$ T = \mu^4 \sim { 1 \over a^4 g^2_{YM} } \Rightarrow \mu^4 \sim { 1 \over L^4_{Planck } ( N g^2_{YM} ) } 
\Rightarrow \mu^{-4}  \sim (N g^2_{YM} ) L^4_{Planck} . \eqno ( 43 ) $$
which has a simliar form as the celebrated Maldacena result relating the size of the $AdS_5$ throat 
$ \rho^4 $ to the 't Hooft coupling $ N g^2_{YM}  $ and the Planck scale $ L^4_{ Planck} \sim (\alpha')^2 $, 
( the inverse string tension squared ). 
In section {\bf 4.3} we shall provide the details supporting these relations.   

Based on the arguments of this section, the generalization of these results to $p$-branes of higher dimensionality , $ p+1 = 2n = 4k$, 
is straightforward and was performed in [34]. 
The starting actions of the Generalized Yang Mills theories GYM ( and their corresponding $\theta$ terms  )  
 were given earlier in eq-(6) following the Lagrangians of Tchrakian [ 10 ] .  
We refer to our work in [34] for all the details that show how the Moyal deformation of these theories, in the 
$quenched$ and reduced  approximation , 
after taking the $ \hbar = 2\pi/N = 0$ limit , yields Dolan-Tchrakian actions 
for spacetime filling $p$  branes , in the conformal gauge, moving in flat $2n$-dim backgrounds.  
The  Chern-Simons $p'$-brane actions are obtained directly from the $\theta$ terms associated with these GYM theories 
and have $nontrivial$ boundary dynamics.

Deformation quantization can also be implemented in the $SU(N)$ Matrix Chern-Simons models defined in $odd$ $D$ dimensions 
and Topological Matrix $BF$ theories [62].  Matrix Chern-Simons models in odd dimensions are defined : 

$$ \epsilon_{\mu_1 \mu_2....\mu_n } Trace~ [ X^{\mu_1} X^{\mu_2}....X^{\mu_n}]. ~~~ X^\mu = N\times N ~matrix.   $$ 
Due to the cyclic property of the trace, and performing a permutation of indices, we require to have $(-1)^{ D -1 } = 1 $ which implies 
that $D$ must be odd or else the action is trivially zero. 

However one cannot take the large $N$ limit any longer using the Moyal deformation procedure of Poisson brackets. 
Instead one has to use the Zariski star product associated with the
deformations of the Nambu-Poisson brackets  ( Jacobians ) [63].  
This is related to the so-called Nambu Quantum Mechanics where the ordinary commutator is replaced by 
the Nambu commutator ( of a product of $n$ matrices , for example )  : 
$$[ A_1, A_2, A_3, ..., A_n ] = A_1A_2..A_n \pm permutations $$. 
The Zariski star product involves also  the deformation 
parameter $ \hbar \sim 1/N $ ( where $N$ is the size of the matrices ) and allows to construct  : 

$$[ {\cal A}_1, {\cal A}_2, {\cal A}_3, ..., {\cal A}_n ]_{\bullet }  = 
{\cal A}_1 \bullet {\cal A}_2\bullet {\cal A}_3 .....\bullet {\cal A}_n \pm permutations $$
and is the analog of Moyal bracket 
whose classical $\hbar = 0$ ( large $N$ ) limit yields the Nambu-Poisson bracket.    
The main result is that a Zariski deformation of these $SU(N)$ Matrix Chern-Simons models yields , 
in the classical $ \hbar = 1/N = 0 $  limit , 
the actions for Chern Simons $p'$-branes, $ p'+ 2 = D$ ( $n=D$ is odd ), using the correspondence 
$ {\cal A} (\sigma ) \leftrightarrow X(\sigma)$ and  after an integration by parts.  
The Topological Matrix $BF$ models , in 
the large $N$ limit, yield the coupling of the dualized Kalb-Ramond field $ {\tilde B}_{\mu_1\mu_2....\mu_{p+1}} $ 
to the $p+1$-dim world volume of $p$-branes.  Or viceversa : the coupling of a Kalb-Ramond field to the 
$dual$ brane characterized by the ${\tilde X} (\sigma )$ variables.

\bigskip 

\centerline {\bf IV} 
\bigskip 

\centerline{\bf 4.1  Gravity from BF-Chern-Simons-Higgs  Theories  }
\bigskip

\bigskip

In the last sections we studied those self-dual $p$-branes moving in ( pseudo ) Euclidean backgrounds related 
to the topological Chern-Simons $p'$-branes when $ p' + 1 = p = D -1$ . We 
found how ( spacetime filling ) $p$-branes moving in flat backgrounds emerge from the large 
$ N = \infty$ limit $\Rightarrow \hbar = 0$ 
of the Moyal deformations of YM and GYM theories in the quenched-reduced  approximation.  
Also we have shown how Chern-Simons branes are directly related to 
the Moyal deformation of the $\theta$ terms associated with these Yang-Mills 
and GYM theories with $nontrivial$ boundary dynamics . 
In this sections we shall study $p$-branes moving in $curved$ $AdS$ backgrounds and later in {\bf 5} the large $N$ limit 
of $SU(N)$ YM in $curved$ backgrounds is studied. In particular YM coupled to gravity and matter ( scalar ) fields.

To begin with we will show how gravitational actions with a cosmological constant can be obtained from 
an action $inspired$ from a  BF-CS-Higgs theory . Our procedure, although very similar in many respects to Wilczek's work [41] , 
$differs$  from his in the $form$ of our action below and it also  
exploits the topological $BF$ orgins of ordinary gravity [40]  and of Yang-Mills theories [46, 48 ].  
It has been known for some time that the first order formalism of pure $3D$ gravity is a $BF$ theory.
The same occurs for $4D$ gravity and higher-dim gravity if additional quadratic constraints on the $B$ field are added to the Lagrangian [40 ] . 
These $BF$ theories are very suitable for the spin-foam quantization techniques [67] that are valid in any dimension due to common structures 
to all  constrained $BF$ theories. A very important result is that $4D$ YM theory can be obtained from a  deformation of a Topological 
$BF$ theory [48]. 

Deformations of a world volume $BF$ theory  as possible deformations of a Topological  open membrane model by 
means of the antifield $BRST$ formalism was performed by Ikeda [46 ].
 Noncommutative structures on the boundaries of the open membrane were obtained as a generalization 
of the path integral representation of the star product deformation given by Kontsevich [ 64 ] . 
The path integral representation of the Kontsevich
formula on a Poisson manifold was given as a perturbative expansion of a two-dim field theory defined 
on the open two-dim disk $D^2$ [ 65 ] . 
Similar star product structures appear in open string theory with a nonzero Neveu-Schwarz $B$ field [ 66 ] .  

Based on the BF-YM relation [48]  , and the branes-YM relation 
of section {\bf 3.2 } we are going to close the triangle BF/YM/Gravity by showing in this section how gravity with a 
negative cosmological constant can be obtained from a BF-CS-Higgs theory. The $4D$ action below is inspired from a BF-CS model 
defined on the boundary  of the $5D$ region $ D^2 \times R^3 $, where $D^2$ is the open domain of the two-dim disk. 
As we said earlier, $AdS_4$ has the topology of $ S^1( time ) \times R^3 $ 
which can be seen as the ( lateral ) boundary of $ D^2 \times R^3 $.

The relevant BF-CS-Higgs $inspired$  action is based on the isometry group of $AdS_4$ space given by 
$SO(3, 2)$ , that also coincides with the conformal group of the $3$-dim boundary of $ AdS_4$ : $ S^1 \times S^2 $.  
The action involves the gauge fields $A^{AB}_\mu $ and a family of Higgs scalars 
$\phi^A~$ that are $SO(3,2)$ vector-valued $0$-forms and the indices run from $ A = 1,2,3,4,5$. 
The action can be written in a compact notation  using differential forms :  

$$S_{BF-CS-Higgs} =  \int_{M_{4}}   \phi \wedge F \wedge F + \phi\wedge  d_A \phi \wedge d_A \phi \wedge d_A \phi \wedge d_A \phi 
 -  V_H (\phi ) . \eqno ( 44 )  $$ 
A word of caution : strictly speaking,  because we are using a covariantized exterior differential $d_A$,  we don't have the standard 
BF-CS theory.  For this reason we use the terminology BF-CS-Higgs inspired model.  
The $5D$ orgins of the BF-CS inspired action is of the form 

$$\int_{D^2 \times R^3}   d\phi \wedge F \wedge F \leftrightarrow \int  B \wedge F_{4} . ~~~ B = d\phi.~~~F_{4} = F \wedge F. $$
$$ \int _{ D^2 \times R^3}  d \phi\wedge d \phi \wedge d \phi \wedge d \phi \wedge d\phi \rightarrow 
\int_{S^1 \times R^3} \phi \wedge d \phi \wedge d \phi \wedge d \phi \wedge d\phi           $$

The $F$ and $ F_{4} = F \wedge F$ fields  satisfy the Bianchi-identity : 

$$ F = d_A A = d A + A \wedge A .~~~d^2_A \phi = F \phi \not=0. ~~~ d^2_A A = d_A F = 0 \Rightarrow  d_A ( F \wedge F ) = 0 $$

The Higgs potential is : 
 
 $$ V_H( \phi ) = \kappa_1 ( \eta_{AB} \phi^a \phi^B - v^2 )^2. ~~~
\eta_{AB} = (+, +, +, -, - ). ~~~\kappa_1 = constant. \eqno ( 45 )  $$ 

The gauge covariant exterior differential is defined : $ d_A = d   + A  $ 
so that   $d_A \phi = d\phi + A \wedge \phi $ and the $SO(3, 2)$ field strengths : 
$ F = dA + A \wedge A $ 
are the usual ones associated with the $SO(3, 2) $ gauge fields in the adjoint representation :  

$$ A^{AB}_\mu = A^{ab}_\mu~ ;~  A^{5a}_\mu ;  ~ a, b = 1,2,3,4. \eqno (46)  $$ 
which, after symmetry breaking,  
will be later identified as the Lorentz spin connection $ \omega^{ab}_\mu $ and the vielbein field respectively : 
$ A^{5a}_\mu = \lambda e^a_\mu $ where $ \lambda $ is the inverse $AdS_4 $ scale.

The Lie algebra $SO(3, 2 ) $ generators obey  the commutation relations : 

$$  [ M_{AB}, M_{CD} ] =  \eta_{BC} M_{AD} - \eta_{AC} M_{BD}  + \eta_{AD} M_{BC} -  \eta_{BD} M_{AC} . \eqno ( 47)            $$

The Higgs potential is minimized  at tree level when the vev are, for example, : 

$$ < \phi^ 5 >  = v. ~~~ < \phi^a >  = 0. ~~~ a = 1, 2, 3, 4. . \eqno ( 48 ) $$ which means that one is 
freezing-in at each spacetime point the internal $ 5$ direction of  the internal space of the group $ SO(3, 2)$. 
Using these conditions (48 ) in the definitions of the gauge covariant derivatives acting on the 
internal $SO(3,2)$-vector-valued  spacetime scalars $\phi^A ( x) $,  we have that at tree level :

$$ \nabla_\mu \phi^5 = \partial_\mu \phi^5 + A^{5a}_\mu \phi^a = 0.   ~~~ 
\nabla_\mu \phi^a = \partial_\mu \phi^a + A^{ab}_\mu \phi^b + A^{a5}_\mu \phi^5 =  A^{a5}_\mu v. \eqno ( 49 )  $$
A variation of the action w.r.t the scalars $\phi^a$ yields the zero Torsion condition after  
imposing the above conditions  (48, 49 ) solely $after$ the variations have  been taken place. 
Varying w.r.t the $\phi^a$ yields the $ SO(3, 2)$-covariantized Euler-Lagrange equations :

$$ {\delta S \over \delta \phi^a } - d_A { \delta S \over \delta ( d_A \phi^a) } = 0 \Rightarrow 
F^{5a}_{\mu\nu}  = T^a_{\mu\nu} = \partial_\mu e^a_\nu  + \omega^{ab}_\mu  e^b_\nu  - \mu \leftrightarrow \nu = 0 
\Rightarrow  \omega^{ab}_\mu  = \omega ( e^a_\mu  ) . \eqno ( 50 ) $$ 
and one recovers the standard Levi-Civita ( spin ) connection in terms of the ( vielbein ) metric .  
A variation w.r.t the remaining $ \phi^5 $ scalar yields after using the relation $ A^{a5}_\mu = \lambda e^a_\mu $ :

$$ F^{ab}_{\mu\nu} F^{cd}_{\rho\tau } \epsilon_{abcd5} \epsilon^{\mu\nu\rho\tau } + 
5 \lambda^4 v^4  e^a_\mu e^b_\nu e^c_\rho e^d_\tau    \epsilon_{abcd5} \epsilon^{\mu\nu\rho\tau } = 0  \Rightarrow $$
$$ - { 1\over 5} \phi^5  F^{ab}_{\mu\nu} F^{cd}_{\rho\tau } \epsilon_{abcd5} \epsilon^{\mu\nu\rho\tau } = 
 \phi^5 \nabla_\mu \phi^a \nabla_\nu \phi^b \nabla_\rho \phi^c \nabla_\tau \phi^d \epsilon_{abcd5} \epsilon^{\mu\nu\rho\tau } \eqno ( 51) $$
these last equations  (48, 49, 50, 51)  will allows us  to eliminate on-shell 
all the scalars $\phi^A $ from the action (44) furnishing the 
MacDowell-Mansouri-Chamseddine-West action  for gravity as a result of an spontaneous 
symmetry breaking of the internal $SO(3, 2)$ gauge symmetry due to the Higgs mechanism 
leaving unbroken the $SO(3, 1)$ Lorentz symmetry : 

$$ S_{MMCW} =  {4 \over 5 }  v \int d^4x ~  F^{ab}_{\mu\nu} F^{cd}_{\rho\tau}   \epsilon_{abcd5} \epsilon^{\mu\nu\rho\tau } . 
\eqno (52)      $$ 
with the main advantage that it is no longer  necessary to impose by hand the 
zero Torsion condition in order to arrive at the 
Einstein-Hilbert action. On the contrary, the zero Torsion condition is a direct result of the spontaneous symmetry breaking 
and the dynamics of the orginal BF-CS inspired action.  
In general, performing  the decomposition 

$$ A^{ab} _\mu = \omega^{ab}_\mu. ~~~ A^{a5} _\mu = \lambda e^a_\mu . \eqno ( 53 ) $$ . 
where $\lambda$ is the inverse length scale of the model ( like the $AdS_4$ scale )  
and inserting these relations 
into the MMCW action yields finally the  Einstein-Hilbert action, the 
cosmological constant plus the Gauss-Bonnet Topological invariant  in $ D = 4 $ :

$$ S = {8\over 5}  \lambda^2 v  \int R \wedge e \wedge e ~+ { 4 \over 5 } \lambda^4 v \int e\wedge e \wedge e \wedge e ~ 
+{ 4 \over 5 }  v \int R \wedge R                    . 
\eqno ( 54 ) $$ 
which implies that the gravitational $L^2_{ Planck } $ 
and cosmological constant $\Lambda_c$ are fixed in terms of $\lambda, v $ , up to numerical factors, as  : 

$$ \lambda^2 v = { 1\over L^2_P} . ~~~ \Lambda_c = \lambda^4 v. \eqno ( 55a) $$
Eliminating the vev value $v$ yields a geometric mean relationship among the three scales : 

$$ \lambda^2 { 1 \over L^2_P } = \Lambda_c 
\Rightarrow L^4_P \leq { 1\over \Lambda_c } \leq { 1 \over  \lambda^4 }~\Rightarrow ~  
{1\over L^4_P }  \geq { \Lambda_c } \geq {\lambda^4 }. \eqno ( 55b)  $$
Hence we have an upper/lower bound on the cosmological constant 
$ \Lambda_c $ based on the Planck scale and the 
$AdS$ inverse scale $ \lambda $.  

We will use precisely these geometric mean relations  (55) to get an estimate of the cosmological constant. 
If one sets the the inverse $AdS_4$ scale $\lambda$ ( the size of the throat )  in terms of the Planck scale $L_P$ : 

$$ \lambda^{ 4} = { 1 \over N L^4_P } $$ 
and inserts it into the geometric mean relations  (55) , one obtains a value for  the cosmological constant : 

$$\Lambda_c = { 1 \over {\sqrt N } } { 1 \over L^4_P } =  { M^4_P  \over {\sqrt N } }. \eqno ( 56)  $$
which is a nice result because in the $ N = \infty$ limit , 
this value of the cosmological constant vanishes ! This corresponds to the flat space limit 
$ \lambda = 0$.  In the next section we will justify setting $ a = ( 1 / \lambda) \rightarrow a^4 = N L^4_P$ 
to derive a similar relation as the Maldacena's expression  
based on the large $N$ limit of $SU(N)$ YM on curved backgrounds ( $AdS_4$ ).

The reason $AdS_4$ is relevant to estimate the cosmological constant ( vacuum energy density ) is because it corresponds to a 
$vacuum$ of the orginal BF-CS-Higgs inspired-action :  
$$ F^{ab}_{\mu\nu} = 0. ~~~ Torsion = F^{a5}_{\mu\nu}  = 0. ~~~ \phi^5 = v. ~~~\phi^a = 0 . \eqno (57) $$
the solutions to (57) incoporate $AdS_4$ spaces in a natural way . 
Using the decomposition of the $SO(3,2)$ gauge fields  (53) in the vacuum equations (57) and eq-( 50 )  
one arrives at :   

$$  F^{ab}_{\mu\nu} = 0 \Rightarrow F^{ab}_{\mu\nu} =  R^{ab}_{\mu\nu } ( \omega (e))  + \lambda^2 e^a_\mu \wedge e^b_\nu  = 0 $$
$$  d \omega + \omega \wedge \omega +   
\lambda^2 e\wedge e= 0 
\Rightarrow R = -    \lambda^2 . \eqno ( 58)  $$  
which is a hallmark of $AdS_4$ space : spaces of constant negative scalar curvature. 
Hence the $ AdS_4$ space is a natural $vacuum$ of the theory associated with the inspired BF-CS-Higgs model (44).  
Based on this fact that $AdS$ spaces are natural vacuum solutions, Vasiliev constructed a theory of massless higher spin fields excitations 
of $AdS_4$ based on higher spin ( higher rank tensors ) algebras whose spin ranges $ s = 2, 3, 4....\infty$ ; i.e 
higher spin massless fields propagating in curved $AdS$ backgrounds. 
This procedure does $not$ work 
in Minkowski spacetime. There is an infinite number of terms in this theory involving arbitrary powers of $\lambda$ . 
This bypasses the no-go theorems of writing consistent interactions of higher spin fields ( greater than $ s = 2 $) in flat Minkowski spacetime.    
\bigskip
\centerline{ \bf 4.2  New types of Couplings of Matter to Gravity and YM fields }
\bigskip

Recently new ( super ) string actions have been given by Guendelmann [56] , that can naturally be 
extended to all $p$-branes where instead of  
using an auxiliary world volume metric, like those used in Polyakov-Howe-Tucker actions to implement reparametrization invariance,   
he introduced a set of $p+1$ auxiliary scalar fields $ \phi^1 (\sigma ), \phi^2 (\sigma ).... $ 
that are coupled to the 
brane coordinates in a nontrivial way through a Jacobian . 
Such Jacobian is made out of the auxiliary scalars 
and plays the role of another measure of integration 
different from using the auxiliary world-volume metric .

$$ J = Jacobian  = \{ \phi^1, \phi^2, ...\phi^n \} \equiv         
{\partial  ( \phi^1, \phi^2, ...\phi^n )  \over  \partial ( \sigma^1, \sigma^2,....\sigma^n )  } . \eqno ( 59)   $$ 
which is also called the Nambu-Poisson bracket. 
In the case of a string $ p+ 1 = 2 $ it becomes the ordinary Poisson bracket. 
Based on the action for a point particle moving in a flat target spacetime : 

$$  S = - { 1 \over 2} \int d\tau e (\tau )  [ - { m^2 {\dot X}^2 \over e (\tau ) ^2  }  + 1 ] . \eqno ( 60)      $$ 
Introducing the auxiliary einbein field $ e = e (\tau ) $ and eliminating  it  from the action via its algebraic equations of motion 
and plugging  its solution back into the original action gives  the well known 
reparametrization invariant action in terms of a ( dimensionless ) 
affine parameter $\tau $ along the particle's worldline : 

$$ S = - m \int d \tau  \sqrt { |{\dot X}^\mu {\dot X}_\mu | } . \eqno (61)     $$ 

Guendelmann generalized such point particle action to the ( super ) string case [56] 
and recently it has been  extended  to all  $p$-branes and Dirac-Born-Infeld actions by us [57]  
after making the following correspondence  in eq-(60) :

$$ (dX/d\tau ) ^2 \leftrightarrow  \{ X^{\mu_1}, X^{\mu_2},. ....., X^{\mu_{p+1}} \}^2. ~~~
m \leftrightarrow  T_p .~~~ \int d\tau ~e \leftrightarrow  \int d^{p+1} \sigma  \{ \phi^1, \phi^2, ...., \phi^{ p+1} \}  $$ 
The new $p$-brane actions are given then : 
$$ S = - { 1\over 2} \int d^{p+1} \sigma ~ \{  \phi^1, \phi^2, ...., \phi^{ p+1} \}  ~ 
[- { T^2_p \over (p+1)! }  { \{ X^{\mu_1}, X^{\mu_2}, X^{\mu_3}, ....., X^{\mu_{p+1}} \}^2  \over 
 \{ \phi^1, \phi^2, ...., \phi^{p+1} \}^2 }  + 1 ]   . \eqno (62 )$$           
such actions are just another version of the reparametrization invariant Schild actions for $p$-branes where the auxiliary field 
$ e $ is replaced in (62)  by the the Jacobian measure involving the auxiliary Guendelmann scalars. Eliminating $e$ from the Schild 
action yields brane Dirac-Nambu-Goto actions.   

To write down new couplings of matter to YM fields , we will use a similar Lagrangian , although $not$ identical ,  to the one  
that  was given earlier by Wilczek [41]
displaying the nontrivial coupling  of the scalars $\phi$'s  to the 
$SU(N)$ Yang Mills field-strength $ G_{\mu\nu} $ . 
One denotes the $SU(N)$ valued  field strength as $ G_{\mu\nu}$  to avoid confusion 
with the previous notation for the $ SO(3, 2)$-valued $F_{\mu\nu} $  .

$$ L_{YM} =  { \kappa  \over g^2_{YM}  }  \epsilon^{\gamma\delta \rho \tau } 
\epsilon^{\alpha \beta \mu \nu  }  
 Tr~ (G^{YM}_{\alpha \beta } G^{YM}_{\gamma \delta } ) \nabla_\mu \phi^A \nabla_\nu \phi^B \nabla_\rho \phi^C \nabla_\tau \phi^D 
 \eta_{AC} \eta_{BD} \eqno ( 63  ) $$
where one must introduce a parameter $\kappa$ of $(length)^4$ dimensions to make the action dimensionless. 

If one wishes to identify the derivative terms as a 
$ SO(3, 2)$-covariantized measure of integration ( a la Guendelmann ) 
we must abandon  
the rerms $ \eta_{AC} \eta_{BD}        
$ used by Wilczek , which are not fully antisymmetric in $all$ of its indices, 
and replace them by a totally 
anti-symmetric tensor $ \Omega_{ABCD} $ made out 
of the the $SO(3, 2)$ structure constants. 
It would be erroneous to use $\epsilon_{ABCD}$ becuase this object breaks the $ SO(3, 2)$ symmetry from the very start.  
Hence selecting a totally anti-symmetric tensor built from the $SO(3, 2)$ structure constants : 

$$ \Omega_{ABCD} = f_{ABEF}^{GH} f_{CDGH}^{EF} - f_{ACEF}^{GH} f_{BDGH}^{EF} + f_{ADEF}^{GH} f_{BCGH}^{EF}....
\eqno (64)                 $$
will furnish the correct terms that must be used in the Lagrangian  by Wilcezk 
in order to recast it in the Guendelmann form . 
The initial choice using the 
$\eta_{AC} \eta_{BD} $ will $not$  work because of the symmetry properties of the $\eta_{AC} = \eta_{CA}$. The structure constants 
are antisymmetric under $A \leftrightarrow B; E \leftrightarrow F$ and also under 
the double exchange $ AB \leftrightarrow EF$.

Having introduced the Lagrangian (63) one can combine it with the three terms in the BF-CS-Higgs action to bring about a 
full BF-CS-Higgs-YM theory .  This is a very rich model per se and the way to proceed is to look at the full four terms of the 
BF-CS-Higgs-YM theory and study this  system as a whole.  
But at the moment we are going to simplify matters and just  
$evaluate$ the latter Lagrangian   ( with the  $\Omega_{ABCD} $ coefficients ) 
in the specific scalar background given by the previous vev 
of the scalars $\phi$ : $ <\phi^5 > = v $ and $ <\phi^a > = 0; ~
a = 1, 2, 3, 4 $ .  
These simplifying conditions fix the gauge covariant derivatives at tree level  :    

$$ \nabla_\mu \phi^a = \partial_\mu \phi^a + A^{ab}_\mu \phi^b + A^{a5}_\mu \phi^5 = A^{a5}_\mu \phi^5 = \lambda v e^a_\mu . 
$$
$$ \nabla_\mu \phi^5 = \partial_\mu \phi^5 + A^{5a}_\mu \phi^a  = 0 .  $$
The vielbeins in curved background satisfy  : 

$$ e^a_\mu e^c_\rho \eta_{ac} \equiv g_{\mu\rho}  . ~~~    e^b_\nu e^d_\tau \eta_{bd} \equiv g_{\nu \tau }  . \eqno (65)  $$ 

The only nonvanishing antisymmetric combinations of the $SO(3,2)$ covariant derivatives  are : 

$$ \nabla_\mu \phi^a \nabla_\nu \phi^b \nabla_\rho \phi^c \nabla_\tau \phi^d \Omega_{abcd} = 
\lambda^4 v^4 e^a_\mu e^b_\nu e^c_\rho e^d_\tau \Omega_{abcd} = \lambda^4 v^4 \Omega_{\mu\nu\rho\tau} ( x ) .  \eqno (66)   $$ 

The $dualized$ YM fields are defined :  
 
$$ {\tilde G}^{\mu\nu} \equiv {1\over 2} \epsilon^{\mu\nu\alpha \beta} G_{\alpha\beta} $$ 
Note that the $covariantized$ permutation symbols are defined : 

$$ e_{\mu\nu\rho\tau } \equiv \sqrt g \epsilon_{\mu\nu\rho\tau}. ~~~~ 
e^{\mu\nu\rho\tau } \equiv { 1 \over \sqrt g }  \epsilon^{\mu\nu\rho\tau}.$$
hence the dualized variables are not strictly speaking the same as the Hodge duals.  
This permits  to rewrite the Guendelmann version of the Wilczek's  action  , 
in the Chern-Simons form  : 

$$  { \kappa \lambda^4 v^4 \over g^2_{YM} } \int d^4 x ~~ e^a_\mu e^b_\nu e^c_\rho e^d_\tau ~\Omega_{abcd} ~ 
{\tilde G}^{\mu\nu}{\tilde G}^{\rho\tau} = 
{ \kappa \lambda^4 v^4 \over g^2_{YM} } \int d^4x~  
\Omega_{\mu\nu\rho\tau} ( x ) ~ {\tilde G}^{\mu\nu}{\tilde G}^{\rho\tau}  \Rightarrow \eqno ( 67 ) $$

$$  S \sim { \kappa \lambda^4 v^4 \over g^2_{YM} } \int d^4 x ~\sqrt {g (x) }
 ~ \epsilon_{\mu\nu\rho\tau } ~{\tilde G}^{\mu\nu}{\tilde G}^{\rho\tau} $$

The Lagrangian (67 ) has as Chern-Simons-like look to it associated with the 
$dualized$ YM fields $ {\tilde G }_{\mu\nu} $ variables after recalling that : 

$$\Omega_{\mu\nu\rho\tau } \equiv  e^a_\mu e^b_\nu e^c_\rho e^d_\tau ~\Omega_{abcd} 
\sim  \epsilon_{\mu\nu\rho\tau} \sqrt {det|g|} $$
The action (67) is  invariant 
under volume-preserving diffs ( it is not fully reparametrization invariant like a Chern-Simons action ).

Barring the subtleties involved in the quenched-reduced approximation in curved backgrounds, as we shall see next, 
the large $N$ limit of the action ( 67 ) has a formal $correspondence$ ,  but is $not$ identical,  
to an action for  a Chern-Simons membrane embedded in curved $4D$ backgrounds . 
The membrane coordinates  are represented in this case by the $dualized$ coordinate variables
$ {\tilde X} (\sigma )  $ , after imposing the  $correspondence$   
$ {\tilde G}^{\mu\nu} \leftrightarrow \{ { \tilde X}^\mu , {\tilde  X}^\nu  \} $ which must $not$ be confused with an actual identity.

For example,  in the large $N$ limit , the volume-preserving diffs invariant Chern-Simons-like looking action ( 67), 
defined in an $AdS_4$ background  , has a formal correspondence ( not an identity ) to a Chern-Simons $ dual $ membrane 
living on the boundary  $S^1 \times S^2$ . The fact that the large $N$ limit involves the $dualized$ variables 
$ {\tilde X}^\mu (\sigma)$ of the Chern-Simons dual membrane is very suggestive within the framework of the $AdS/CFT$ duality correspondence.

Had we used Wilczek's initial Lagrangian from the very start, without using the appropriate Guendelmann measure,  
one would have obtained an expression : 
$$  S =          { \kappa \lambda^4 v^4 \over g^2_{YM} } \int d^4x~  G^{\mu\nu} G_{\mu\nu} \eqno ( 68 ) $$
invariant under volume-preserving diffs.  Eq-(68) is obtained after using the identities 
$ {\tilde G }^{\mu\nu} {\tilde G  }_{\mu\nu}  = G^{\mu\nu} G_{\mu\nu}$ and the relations ( 65 ) which define the metric tensor.   
It is at this point where we will encounter technical  subtleties if one naively uses the large $N$ quenched-reduced approximation 
to the $SU(N)$ YM  theory in $curved$ backgrounds. 
To contract spacetime indices in the action (68) requires the spacetime metric , which in turn, has a nontrivial world-volume 
dependence : $ g_{\mu\nu} (   X(\sigma)      ) $.    

\bigskip

\centerline {\bf V.  The large $N$ limit of $SU(N)$ YM in curved backgrounds } 
\bigskip 

In { \bf 3.2 } we investigated the large $N$ limit of quenched QCD in $flat$ backgrounds which generated brane actions ( Chern-Simons branes 
from the $\theta$ terms ).   
Another interesting question is what is the large $N$ limit of $SU(N)$ YM in $curved$ backgrounds. 
We will see that this limit is connected to $W$ Geometry. We have shown in [22 ] that 
$SU(N)$ Self Dual YM in flat $4D$ spaces, in the large $N = \infty $ limit, was given by an effective $ D = 6$-dim theory. 
A dimensional reduction from $ D = 6 \rightarrow D = 4$ that involved a mixing of " colour " degrees of freedom and spatial ones 
( mixing of colour and spacetime coordinates  )  furnished Plebanski equations for 
$ 4D$ Self Dual  Gravity. If one starts already with a $curved$ background the large $N$ limit of $SU(N)$ SDYM is going to be 
more involved because one cannot longer mix the $N = \infty$ colour indices ( $\sigma$ coordinates ) 
 with spacetime variables in a trivial way. This will be the purpose of this section.

The YM action for a $SU(N)$ YM theory in a curved background ( say $AdS_4$ ) :

$$S = - { 1\over g^2_{YM} } \int d^4 x \sqrt g ~ G_{\mu\nu}G^{\mu\nu} . \eqno ( 69 ) $$ 
Suppose we wanted to proceed along identical lines as section {\bf 3.2 } to perform the large $N$ limit via a Moyal 
deformation procedure. This is the moment where we are going to encounter ambiguities and problems in using 
the quenched, reduced approximation  of section {\bf 3.2}.  

Firstly, due to the presence of the density $\sqrt {g (x)}   $ that varies from point to point, 
in (68) one will end up by having a $variable$ tension. 
As  one changes the reference point $x_o$, one can reabsorb $\sqrt g $ in the definition of the tension.   
Since the background spaces are no longer $flat$ , quenching and reducing the theory 
to a single point loses its invariant meaning. Flat spaces are isotropic and homogoneous and the 
translation operator is indeed an 
invariant Casimir operator of the algebra. This is not the case in $AdS$ spaces .  
The $P_\mu$ operator is not a Casimir of the 
$ SO(3, 2 ) $ algebra. Its obstruction is the curvature. 

Despite these subtleties, and the problem of having a $variable$  tension, let us suppose for a moment that we were to use 
the quenched-reduced approximation ( in flat spaces ) of section {\bf 3.2 } to the action (68) 
that generated branes in the large $N$ limit when the background spaces were flat. 
What is going to happen now is that we will encounter three different possibilities 
and only one of them has the correct form of a Dolan-Tchakian action for a $3$-brane action on curved $AdS_4$ : . 

Because one has now the three combinations : 

$$ G^{\mu\nu} G_{\mu\nu} = G^{\mu\nu} G^{\rho\tau } g_{\mu \rho} g_{\nu\tau } = 
G_{\mu\nu} G_{\rho\tau } g^{\mu \rho} g^{\nu\tau }. \eqno ( 70 ) $$
one will generate three different expressions in the large $N$ limit. 
The product of the $SU(N) $ YM field strengths becomes , after using the 
WWGM deformation quantization procedure that maps products 
of matrix-valued operators into the Moyal star products of their corresponding symbols  
${\cal A}_\mu ( q^i, p^i ) = {\cal A_\mu} ( \sigma )$  :

$$ G^{YM}_{\alpha \beta } G^{YM}_{\gamma \delta } \Rightarrow  
\{ {\cal A}_{\alpha} , {\cal A}_{\beta } \} *  \{ {\cal A}_\gamma , {\cal A}_\delta \} \Rightarrow 
\{ X_{\alpha} , X_{\beta } \} *  \{ X_\gamma , X_\delta \} . 
\eqno (71) $$

Finally, using eqs-(68, 70 ), after taking  the classical limit $ \hbar = 0 $ , 
which is equivalent to taking the large $ N $ limit, 
turns Moyal brackets into Poisson brackets and noncommutative star products into ordinary ones ,  the large $N$ limit 
the YM action becomes : 

$$   {G}_{\mu\nu} {G}_{\rho\tau} g^{\mu\rho} g^{\nu \tau } . ~~~ N = \infty \Rightarrow $$
$$ S = - ( { 2 \pi \over a}    )^4 { 1\over g^2_{YM}  }\int d^4 \sigma ~ \{ X_\mu, X_\nu \}_{PB} 
\{ X_\rho ,   X_\tau  \}_{PB}~ g^{\mu \rho  } g^{\nu\tau    }.    \eqno (72 ) $$

It is at this point we encounter a $caveat$ . Before one tries to relate the action (72),  
in a naive fashion,  to the Dolan-Tchrakian  
action ( in the conformal gauge )  for a $3$-brane we must stress that the action (72) 
would be solely equivalent to the Dolan-Tchrakian action ( in the conformal gauge ) 
if the embedding four-dim background were $flat$. Because the background is $curved$ 
one cannot naively raise and lower spacetime indices $inside$ the Poisson brackets : 
one $cannot$  naively set : 
$ \{ g_{\mu\rho } X^\rho , g_{\nu \tau} X^\tau \} \equiv   \{ X_\mu, X_\nu \} $        
to be equal to : $g_{\mu\rho} g_{\nu\tau} \{ X^\rho,  X^\tau \} $ 
as required in the Dolan-Tchrakian action . The reason is that now the Poisson brackets involving  
the $AdS_4$ background metric components  
$ \{ g_{\mu \rho } ( X (\sigma ) ), X^\tau (\sigma ) \} $ are clearly $ not ~ zero$. 
Not even in the conformally flat case.

Thus the action (72) is $not$ equivalent to the Dolan-Tchakian action for a $3$-brane moving in a curved 
$AdS_4$ spacetime ( in the conformal gauge ). 
In order to arrive at the Dolan-Tchrakian action one needs the choose the gauge/fields cosrrespondence in a different way 
that takes into account the subtleties of the index-position  : 

 $$ G^{\alpha \beta } G^{\gamma \delta } \Rightarrow  
\{ {\cal A}^{\alpha} , {\cal A}^{\beta } \} *  \{ {\cal A}^\gamma , {\cal A}^\delta \} \Rightarrow 
\{ X^{\alpha} , X^{\beta } \} *  \{ X^\gamma , X^ \delta \}. ~~~ N = \infty \Rightarrow  
\eqno ( 73a  ) $$

$$ S = - ( { 2\pi \over a}   )^4  { 1 \over g^2_{YM}  }\int d^4 \sigma \{ X^\mu, X^\nu \}_{PB} 
\{ X^\rho ,   X^\tau  \}_{PB}~ g_{\mu \rho  } g_{\nu\tau    }.    \eqno (73b ) $$
which has now has indeed the correct Dolan-Tchrakian form in $curved$ backgrounds. The remaining hurdle lies now in the 
presence of the density $ \sqrt { g ( x) } $ , which could be reabsorbed into a $variable$  tension.

The third possibility yields another expression that $differs$ from the previous two :

$$  S = - ( { 2\pi \over a}   )^4  { 1 \over g^2_{YM}  }\int d^4 \sigma \{ X^\mu, X^\nu \}_{PB} 
\{ X_\mu  ,   X_\nu  \}_{PB}. \eqno (74 ) $$

Therefore, we have encounter definite ambiguities  and problems 
in using the quenched-reduced approximation to evaluate 
the large $N$ limits of actions of $SU(N)$ YM in $curved$ backgrounds. 
There is no ambiguity whatsoever in $flat$  backgrounds, as the results in section 
{\bf 3.2  } have shown. In essence what this means is that the 
Dolan-Tchakrian actions when $ D = 2n $ no longer have the Yang-Mills form in curved backgrounds $ F \wedge ^* F $.

The correct way to proceed  in order to avoid these ambiguities,  due to the nature of curved spaces,  
is to write the $precise$  expression 
that  does $not$  involve quenching nor reduction. 
We will see now why the large $N$ limit of $SU(N)$ YM in $curved$  backgrounds differs from the results in $flat$ backgrounds and 
is related to the gauge theories of $w_\infty, w_\infty^\infty$ algebras  which are the infinite-dim algebras  
of area preserving diffs in two  and four dimensions , respectively.

Similar results will follow for the actions  of Generalized Yang Mills ( GYM)  
theories in $curved$ $4k$-dim backgrounds . A  deformation of the $ SO(4k)$ Lie algebra associated with these 
GYM theories in $curved$ spaces yields  Generalized Gravitational theories [10] in $ D = 4k + 4k  $-dimensions. 
GYM theories are defined by  Lagrangians of the form $ ( F \wedge F \wedge ....\wedge F)^2 $ in $ 4k$-dimensions 
and Generalized Gravitational theories are defined by Lagrangians of the type : 
$ ( R \wedge R \wedge ....\wedge R)^2 $ in $ 4k$-dimensions. The wedge products involve $k$-factors. $ F$ and $R$ are two-forms.

We shall begin first  with the old results which allow to write the correct large $N$ limit of $SU(N)$ YM defined in a $4D$ curved manifold 
$\Sigma$ .  We will based our findings in the results of [37-39] based on the construction of gauge theories of the Virasoro 
and higher conformal spin $w_\infty, w_\infty^\infty$ algebras.  
Vasiliev's construction of higher spin theories on $ AdS$ spaces [ 15] is analogous to the Fedosov deformation 
quantization [18] by using, instead, auxiliary spinorial coordinates for the " fermionic phase space  " . 
The higher spin symmetries are realized by the algebras of oscillators carrying spinorial representations of the spacetime symmetries.  
The conformal group in $ 3D$ is also the isometry group of $AdS_4$ : $SO(3,2)$ and whose algebra is locally 
isomorphic to the symplectic algebra in four-dimensions : $ so(3, 2) \sim sp(4, R)$ . 
Hence, it is not surprisng that symplectic geometry has an important role in all these models.

Park [  36 ]  had already made these result manifest when he showed that 
$4D$ Self Dual Gravity is in fact related to a WZNW model based on $SU(\infty)$. 
$4D$ SD Gravity is  the  non-linear sigma model based in $ 2D $ 
whose target space is  the " group manifold "  of area-preserving diffs of another $ 2D $-dim manifold. 
Roughly speaking, this is saying that the effective $ D = 4 $ manifold where 
Self Dual gravity is defined  is " spliced " into 
two $2D$-submanifolds : one submanifold is the orginal $2D$ base manifold
where the non-linear sigma model is defined. The other $2D$ submanifold is the target group manifold 
of area-preserving diffs of a two-dim sphere $S^2$ .   

Cho et al [37 ] went further and generalized this particular Self Dual Gravity case  
to the full fledged gravity in $ D =  2 + 2 = 4 $ dimensions, 
and in general,  to $any$ combinations of $ m+n$-dimensions. 
Their  main result is that $m+n$-dim Einstein gravity can be identified with 
an $m$-dimensional generally invariant gauge theory of $ Diffs ~ N$ , where $N$ is an $n$-dim manifold. 
Locally the $m+n$-dim space can 
be written as $ \Sigma = M \times N $ and the metric $g_{AB}$ decomposes :

$$ g_{AB} \Rightarrow  \gamma_{\mu\nu} ( x, y ) + e^2 \phi_{ab} ( x, y ) A^a_\mu ( x, y ) A^b_{\nu} ( x, y ) . ~~~ 
e A^a_\mu ( x, y ) \phi_{ab} ( x, y ). ~~~
\phi_{ab} ( x, y ) .... \eqno (75) $$
where $e$ is the gauge coupling constant. This decomposition must $not$ be confused 
with the Kaluza-Klein reduction where one imposes an 
isometry  restriction on the $g_{AB}$  
that turns $A^a_\mu$ into a gauge connection associated with the gauge group $G$ generated by isometry. 
Dropping the isometry restrictions allows $all$  the 
fields to depend on $all$ the coordinates $x, y $. 
Nevertheless $ A^a_\mu ( x, y )$ can still be identified as a connection associated with the infinite-dim gauge group 
of $Diffs~ N$ . The gauge transformations are now given in terms of Lie-brackets and Lie derivatives :

$$ \delta A^a_\mu = - { 1\over e } D_\mu \xi^a = 
- { 1\over e } ( \partial_\mu \xi^a - e [ A_\mu, \xi ]^a ) = - { 1 \over e } 
(  \partial_\mu - e {\cal L}_{A_\mu } )   \xi^a $$
$$ A_\mu \equiv  A^a_\mu \partial_a. ~~~ 
{\cal L}_{ A_\mu } \xi^a \equiv   [ A_\mu, \xi ]^a $$

$$\delta \phi_{ab} = - [ \xi , \phi ]_{ab} . ~~~ \delta \gamma_{\mu\nu} = - [ \xi, \gamma_{\mu\nu } ] . \eqno (76) $$

In particular, if the relevant algebra is the area-preserving diffs of $S^2 $ , given by the suitable basis dependent limit 
$ SU(\infty)$ ,  one induces a natural Lie-Poisson structure generated by 
the gauge fields $ A_\mu $. The Lie derivative of $f$ along a vector $\xi$ is the Lie bracket $ [ \xi, f ] $ which coincides 
in this case with the Poisson bracket $\{ \xi, f \}$.
This implies that  the Lie brackets of two generators of the area-preserving diffs $S^2$ is given precisely by the generator 
associated with their respective Poisson brackets  ( a Lie-Poisson structure ) : 

$$ [ L_f , L_g ] = L _{  \{f, g   \} }.  \eqno ( 77) $$
This relation is derived by taking the vectors $\xi_1^a, \xi_2^a$, along which we compute the Lie derivatives, to be 
the symplectic gradients of two functions $ f(\sigma^1, \sigma^2), g(\sigma^1, \sigma^2 ) $ : 

$$ \xi^a_1 = \Omega^{ab}\partial_b f. ~~~  \xi^a_2 = \Omega^{ab}\partial_b g $$
When nontrivial topologies are involved one must include harmonic forms $\omega$ into the 
definition of $\xi^a$ [42] allowing central terms for the algebras. 
This relation can be extended to the volume-preserving diffs of $N$ by means of the Nambu-Poisson brackets : 

$$ \{ A_1, A_2, A_3,......A_n \} = Jacobian = 
{ \partial ( A_1, A_2, A_3,....., A_n ) \over \partial ( \sigma^1, \sigma^2, ....\sigma^n ) }  \Rightarrow $$
$$[ L_{ A_{1} } , L_{A_{2} }, .........., L_{A_{n} }     ] = L_{  \{ A_{1} , A_{2},........., A_{n}   \} } . \eqno ( 78 ) $$
which states that the Nambu-commutator of  $n$- generators of the volume-preserving diffs of ${\cal N} $ 
is given by the generator associated with their corresponding Nambu-Poisson brackets.  
The generators are obtained in this case by taking the multi-symplectic gradients of functions 
$f_1, f_2....f_{n-1} $ of $\sigma^1, \sigma^2....\sigma^n$ given in terms of the inverse of the multi-symplectic $n$-form $\Omega$  :

$$\xi_{(i)}^{a_1} = \Omega^{a_1 a_2 a_3....a_n} \partial_{a_2}f^{(i)}_1 \partial_{a_3}f^{(i)}_2....\partial_{a_n}f^{ (i) }_{n-1} .....$$ 

When the dimension of ${\cal N} $ is even , locally one can write the volume form in terms of products of area-forms 
$\Omega = \omega \wedge \omega....$ ; i.e  in an appropriate frame the Jacobian ( Nambu-Poisson bracket ) factorizes locally 
into a product of ordinary Poisson brackets. However this doesn't mean that area-preserving has a one-to-one correspondence with 
volume preservings. There are volume-preserving diffs that do not necessarily preserve areas.  
The suitable  star product will be the Zariski product related to the deformation program of Nambu-Poisson mechanics ; i.e. 
deformation theory in multi-symplectic geometry.

The curvature scalar $R^{ ( m+n ) } $ decomposes into :

$$ - { 1 \over 16 \pi G} R^{ ( m+n) }  \Rightarrow  - { 1 \over 16 \pi G} \sqrt{ det |\gamma_{\mu\nu}| }  
~\sqrt { det ~|\phi_{ab}| } \times \{ 
~ \gamma^{\mu\nu} R^{(m)}_{\mu\nu } + { e^2 \over 4 } \phi_{ab} F^a_{\mu\nu} F^{b}_{\rho\tau } \gamma^{\mu\rho} \gamma^{\nu\tau } 
+ $$ 
$$ \phi^{ab}R^{ (n)}_{ab}  +  { 1\over 4} \gamma^{\mu\nu} \phi^{ab} \phi^{cd} D_\mu \phi_{ab} D_\nu \phi_{cd } +....
{ 1\over 4} \phi^{ab} \gamma^{\mu\nu} \gamma^{\rho\tau} 
\partial_a \gamma_{\mu\nu} \partial_b \gamma_{\rho\tau } +..... \nabla_\mu J^\mu +\nabla_a J^a \} \eqno ( 79 ) $$

Therefore, Einstein gravity in $m+n$-dim describes an $m$-dim generally invariant field theory under 
the gauge transformations or $Diffs~N$ . Notice 
how  $A^a_\mu$ couples to the graviton $\gamma_{\mu\nu} $ , 
meaning that the graviton is 
charged /gauged  in this theory and to the $\phi_{ab}$ fields. 
A charged gravity may be related to the recently constructed Nonabelian Geometry [  59 ] . 
The metric $\phi_{ab}$ on $N$ can be identified as a non-linear sigma field whose self interaction potential term is given by :
$\phi^{ab}R_{ab}^{ (n)}$.  
The currents $J^\mu, J^a$ are functions of $\gamma_{\mu\nu}, A_\mu, \phi_{ab}$.  
Their contribution to the action is essential when there are boundaries involved; i.e. the $AdS/CFT$ correspondence.

When the internal manifold ${\cal N}$ is a homogeneous compact space one can perform a harmonic 
expansion of the fields w.r.t the internal $y$ coordinates, and after integrating w.r.t these coordinates $y$, 
one will generate an infinite-component field 
theory on the $m$-dimensional space.  
A reduction of the $Diffs~ N$, via the inner automorphims of a subgroup  $G$ of the $Diffs~ N$,   
yields  the usual Einstein-Yang-Mills theory interacting with a nonlinear sigma field.  
But in general, the theory described in ( 81 ) is by far richer than the latter theory.

A crucial fact of the decomposition ( 77 ) 
is that $each$ single term in (  81 ) is by itself independently invariant under $Diffs~ N$. 
The second term , for example, :

$$ { 1 \over 16 \pi G} \sqrt{ det |\gamma_{\mu\nu}| }  
~\sqrt { det ~|\phi_{ab}| } \times { e^2 \over 4 } \phi_{ab} F^a_{\mu\nu} F^{b}_{\rho\tau } \gamma^{\mu\rho} \gamma^{\nu\tau } . \eqno ( 80) $$
is precisely the one that will be related to the large $N$ limit of $SU(N)$ YM on $curved$  backgrounds. 

One can  choose a preferred-volume  gauge in the internal space ${\cal N} $ so that $ \sqrt {det |\phi_{ab} | } = 1 $ 
leaving a $residual$ symmetry under volume-preserving diffs $N$  . 
Identifying  the coordinates $y^1, y^2...y^4 $ of ${\cal N} $ 
with  the $4$  coordinates  $\sigma^1, \sigma^2...\sigma^4 $ of a " world volume of a $3$-brane  sitting 
"  at each spacetime point $x$  ; 
using the decomposition of a four-volume form $\Omega$ in terms of 
the exterior product of the area-two-forms , $\Omega_{ (4)  } = \omega \wedge \omega $ ;  
identifying the $ \gamma_{\mu\nu } $ with $g_{\mu\nu }$ 
and recurring to the Lie-Poisson structure of eq- ( 79 ) 
that states that the Lie bracket of two generators of the area-preserving diffs 
is given by the generator associated with their Poisson brackets 
will allow  us to impose the  one-to-one correspondence between :

$$  F^a_{\mu\nu }\partial_a  = [ D_\mu, D_\nu ]^a\partial_a   \leftrightarrow 
{\cal F}_{\mu\nu}  ( x, \sigma ) \equiv  \partial_{[\mu}  {\cal A}_{\nu]}  ( x, \sigma ) 
+ \{ {\cal A}_\mu ( x, \sigma ), {\cal A}_\nu ( x, \sigma ) \}_{PB}  . \eqno (81a ) $$ 
so that  when   $ \sqrt {det |\phi_{ab} |}  = 1 $ :

$$\int d^4\sigma ~ \phi_{ab} ~F^a_{\mu\nu} F^{b,\mu\nu  } \leftrightarrow  
\int d^4 \sigma [{\cal F}_{\mu\nu}  ( x, \sigma ) {\cal F}^{\mu\nu}  ( x, \sigma )] .  \eqno ( 81b) $$

There is however a $mismatch$ in (81b) because the l.h.s involves the metric $\phi_{ab}$ components of the internal 
space ${\cal N}$ which are clearly $not$ constant.  To remedy this problem, 
one must work with the most general $curved$ phase spaces ( symplectic manifolds ) ; i.e. 
when the world-volume of the " branes sitting " at each spacetime point $x^\mu $,  are themselves $curved$.  
The naive Moyal star product ceases to be $associative$ in  curved phase space and the Fedosov $\bullet$ product and trace is essential [18] .  
In this case the effective gravitational theory in the $8D$ space,  which locally can be witten as $ \Sigma \times {\cal N} $ , 
is connected to gravity in the $cotangent~ bundle $ of a symplectic four manifold $\Sigma$.  

Therefore , $\Sigma$ must be a symplectic four-manifold itself to carry out the one-to-one correspondence (  81 ) . 
Hull constructed $2D$  $w_\infty$ gravity from a $4D$ Self Dual Gravity associated with  the cotangent space of a 
$2D$ Riemann surface [43].  The author [18] has shown that Hull's construction is in 
fact related to a Fedosov deformation program in curved phase spaces .   
Hence it is the Fedosov trace that must be invoked in eq-( 81 ) to attain a proper correspondence. 
The Fedosov trace is very elaborate,  and recaptures the role of using the internal metric $\phi_{ab}$ to contract indices. 
These results will be reported in  [71] . 

Hence, the correct correspondence , in the large $N$ limit,  is then : 

$$\int d^4\sigma ~ \phi_{ab} ~F^a_{\mu\nu} F^{b,\mu\nu  } \leftrightarrow 
lim ( \hbar = 0 ) ~Tr_{Fedosov}~ [ {\cal W}_{\mu\nu}  ( x, \sigma ) \bullet {\cal W}^{\mu\nu}  ( x, \sigma ) ] . \eqno (81c) $$
where the Weyl-algebra-bundle-valued field :

$$ {\cal W} = d {\cal A} + {\cal A} \bullet \wedge {\cal A}$$ 
must be used, see [ 18 ].

In the limiting $flat$ phase space case, the Fedosov trace reduces to 
the standard one used in {\bf 3.2} given in terms of the ordinary Moyal brackets :

$$ Tr_{Fedosov}~ {\cal W}_{\mu\nu}  ( x, \sigma ) \bullet {\cal W}^{\mu\nu}  ( x, \sigma ) \rightarrow \int d^4 \sigma 
{\cal F}_{\mu\nu}  ( x, \sigma ) *  {\cal F}^{\mu\nu}  ( x, \sigma )  $$
and , in the consequent $\hbar = 0 $ limit , we will then have a matching with the l.h.s of ( 81b ) , as it ought to be, because in this
special case the metric components $\phi_{ab}$ are constant. 
In the more general curved phase space scenario we have to use the rather intricate expressions for 
the Fedosov Trace and the $\phi_{ab}$ components will $no$  longer be trivially constant. 
Matrix models of the type $ \Phi [X, Y ]^2  $ have been investigated in the past   
and in the large $N$ limit,  have a direct matching with the l.h.s of ( 81b ).

To sum up :  Fedosov's deformation quantization is needed to show that the large $N$ limit of $SU(N)$ YM on a $curved$  
four-dim  symplectic space $\Sigma$ can be subsumed by a ( restricted ) 
Einstein gravitational theory  in $ D = 8 $ dimensions, the cotangent bundle of $\Sigma_4$. 
Locally the $ D = 8 $ space can be written as 
$\Sigma_4 \times {\cal N} $. The internal space ${\cal N} $ can be seen as the  $4$-dim  " world-volume " of a $3$-brane  
sitting at each spacetime point $x^\mu$ of $\Sigma$.

Notice that the results of Cho et al [ 37] are valid in general and do not require  
having to use cotangent spaces to symplectic manifolds.  In particular, 
$4D$ Gauge theories based on the Virasoro algebra [38 ] ( diffs of a circle ) 
and the $w_\infty$ ( area-preserving diffs )  algebras have been constructed in [ 39] using the Feigin-Fuks-Kaplansky representation.  
They share the same physical interpretation. Because $w_\infty$ is the algebra of area-preserving diffs in 
$two$-dimensions one must now have for the internal space a 
$two$-dimensional one, and the appropriate action then is described by an effective $ D = 6 $-dimensional action 
Integrating  afterwards w.r.t the two internal variables yields an effective $4D$ spacetime  
theory with an infinite numbe of field-components.  Such actions have been constructed 
in full detail in [ 38, 39].  Moreover,  Higgs matter fields were introduced with a typical quartic potential and which 
generated  towers of massive spin $2$ fields  ( massive higher spin fields  in the case of $w_\infty$ gauge theories  ) 
after an sponteaneous symmetry breaking.  
Massless fields in $AdS$ can be obtained by taking products of two singletons. Products of 
three or more singletons yield massive fields corresponding to the massive representations of $AdS$ groups.   
In section {\bf 5 } we'll see the importance that these gauge theories of the Virasoro and $w_\infty, w_\infty^\infty $ algebras 
have in the  $W_\infty$ Geometry.

Another interesting proyect that warrants a further investigation is to study branes moving on product manifolds of the type 
$ M \times G$ , where $ G$ is group manifold associated with the $volume$ preserving diffs of 
another manifold .  Mason and Sparling have shown that $4D$ Self Dual Gravity can be obtained 
from the Self Dual equations associated with a gauge theory 
of volume preserving diffs in four dimensions , after a dimensional reduction from the effective 
$ D = 8 $ theory down to $ D = 4 $ is performed. The $4D$ Plebanski's equations are recovered also in this way.  
Also relevant is to study $p$-branes moving on $M \times {\cal L} {\cal G} $ where the latter  is the Loop Space  
associated with the maps of a circle $ S^1 $ onto a Lie group $G$. 
This is the familiar case described by ( untwisted affine ) Kac-Moody-Virasoro  algebras. 
Loop algebras associated with $w_\infty$ algebras have been discussed in [42]. 
The generators are functions of the circle coordinate 
$ \sigma $ ( an angle ). A Fourier mode decomposition followed by an integration w.r.t the angle coordinate 
allows to construct similar types of actions like those given in [37-39] based on gauge theories of the Virasoro and $w_\infty$ algebras.

To summarize the contents of {\bf 5.1} : 

Using the main result of Cho et al [37] that gravity in $m+n$ dimensions can be identified to an $m$-dimensional 
generally invariant gauge theories  of $ Diffs ~{\cal N} $ and the crucial role of 
Fedosov's deformation quantization in curved symplectic phase spaces one has  :

( i ) The large $N$ limit of $SU(N)$ YM in a $curved$  symplectic space $\Sigma_4$ can be subsumed inside 
a ( restricted ) $8D$ gravitational theory corresponding to the cotangent bundle of $\Sigma_4$.    
The one-to-one correspondence between the classical $\hbar = 0$  ( large $N$  ) limit of the Fedosov deformation program 
and the gauge theories of area-preserving diffs of a  $4D$  curved internal ${\cal N}$ space,  were given by eqs-( 81c ) 
The full details of this result is currently under investigation [71 ] .  

(ii) Nonlinear sigma models in $ m$-dimensions  whose target spaces are the volume-preserving diffs of a 
$ n$-dim manifold are related to a restricted gravitational theory in $m+n$ dimensions. 
Generalized Self Dual Gravity in $4k$-dimensions [10]  are the higher  dimensional extension of Park's results [36].  

(iii ) The deformation quantization program of the $4k$-dim Generalized Yang-Mills in $curved$  backgrounds 10] 
should be given by an effective Generalized Gravitational theory in $8k$-dimensions [ 10] . 
The introduction of the $4k$- auxiliary coordinates stems  
from the Fedosov deformation quantization of the symplectic structures which emerge from the large 
$N$ limit associated with GYM theories.      

(iv) The more general theories of volume-preserving diffs in $odd$ dimensions 
require the Zariski star product , related to the deformation quantization of   
Nambu-Poisson brackets. We are not aware of what is the analog of the Fedosov deformation program in 
the  odd-dimensional " curved multi-symplectic "  geometry  case . 
 
( v  ) the $zero~ mode~ sector$  of the large $N$ limit of $SU(N)$ YM theories in $curved$ backgrounds   
yields  $3$-branes Dolan-Tchrakian actions ( in the conformal gauge ). 

\bigskip

\centerline { \bf 5.2 A Plausible Geometrical Interpretation of the $AdS/CFT$ duality } 

\bigskip 

After this long route through section {\bf 5.1} we are finally in a position to derive the analog of the 
Maldacena relation (43 ) from the results of the previous section and provide with a very plausible 
geometrical  interpretation of the Maldacena's $AdS/CFT$ correspondence. Since the $correct$ large 
$N$ limit of $SU(N)$ YM in curved symplectic spaces can be subsumed by an effective $ m+m $-dimensional Gravitational  
Theory,  associated with the cotangent space of the symplectic manifold, it is perfectly legitimate to 
introduce now the Planck Scale into the whole picture $ L_P$.    

Therefore it is perfectly legitimate to write : $ \lambda^{ -4} = N L^4_P$ in eq-(55, 56 ) and obtain a  
very appealing value of the cosmological constant : 
$ \Lambda_c = M^4_P/ \sqrt N $  ( which becomes small in the large $N$ limit )  
and to impose the condition on the scale $ a = ( 1/\lambda)$ in eq-(43) 
( $ a$ was the inverse lattice spacing of the large $N$ QCD that generated bags ) 
asociated with the world volume of a bag  sitting at each spacetime point $x$.  
This  resulted in an expression for ( 43) similar to Maldacena's result  $\rho^4( AdS_5) \sim N g^2_{YM} L^4_P$.  We 
believe this is $no$ numerical coincidence but stems from the fact that the 
underlying theory , when the large $N$ limit is taken, is higher-dimensional gravity as the results of [37] 
indicated. For the role that the holographic renormalization group has in the cosmological constant, we refer to [ 68].  

The results of Cho et al [37]  are valid in the more general case than cotangent spaces to symplectic manifolds . 
For instance, let us take $AdS_4$ gravity . 
The  topology  of $AdS_4$ is  $S^1 \times R^3$ . 
The results of Cho et al [ 37 ] imply the following . Lets take the two $dual$ cases : 

(i) $ m = 1, n = 3 $ and (ii) $m = 3, n =1$.  In the first case ( i) we have a generally invariant gauge theory 
of diffs $R^3$ defined over the circle $S^1$. In the second case we have 
a gauge theory of diffs $S^1$ defined over $R^3$.  
This is the gauge theory of the Virasoro group ; i.e the gauge theory of an internal string.  Not surprising, 
conformal invariance is an essential  ingredient. 

The crux of the argument is to view $AdS_4$ gravity from the boundary $S^1 \times S^2$ perspective.  
Since $AdS_4$ is a homogeneous hyperbolic space of constant negative scalar curvature ,  
we have then  two " dual "  perspectives  of $AdS_4$ gravity from the $3D$ boundary  : 

 $$ ( i) a~ gauge~theory~ of~ diffs~ S^1  ~defined~on ~S^2   $$

$$ ( ii) a~gauge~theory~ of~ diffs~ S^2  ~defined~on ~S^1   $$

Setting asides issues of signature , it is very reasonable 
to suggest that  : these two gauge theories are not only two dual perspectives of the same 
$AdS_4$ gravitational theory,  viewed from the boundary, but also they are two dual theories in the sense of 
the srong/weak coupling duality  : $ g^2_1 g^2_2 \sim 1 $.     
The area-preserving diffs $S^2 $ is isomorphic to $SU(\infty)$, hence the large $N$ limit of $SU(N)$ YM appears very naturally. 

In the next section  we'll  discuss noncritical $W_\infty$ strings moving in $AdS_4 \times S^7$ ; their relation to the physics of 
Self Dual Membranes living on the boundary $S^1 \times S^2$ , and to the $3D$ continuos $SU(\infty) (SL(\infty))  $ Toda theory, 
that has an underlying $W_\infty$ symmetry.  We will see once more how gauge theories of 
area-preserving diffs ( like $w_\infty, SU(\infty)$ ) are  precisely neededed to establish :

$\bullet$ The one-to-one correspondence between higher $conformal$ spin symmetries in $3D$ , 
an $extended$  CFT having a $W_\infty$ symmetry $ \leftrightarrow $ 
Vasiliev's higher spin symmetries in the $target$ space given by $AdS_4$.  Vasiliev's construction is based on deformations of the 
algebra $so(3, 2 ) \sim sp (4, R ) $  which is the algebra related to the conformal group on the $3D$ boundary $S^1 \times S^2$ 
and is also the isometry group of $AdS_4$.  
Such $3D$ theory with a higher conformal spin symmetry is precisely related to the noncritcal 
$W_\infty$ string which is an $effective$ $3D$ theory . It is a theory of a " thickened" string ; it is a membrane-like theory. 
We recall that when one takes the $ N =\infty$ limit new dimensions are induced.  
Hence $W_\infty$ strings are effectively a $3D$ extended CFT.

\bigskip

\centerline{ \bf VI} 
\bigskip

\centerline { \bf 6.1  $W$  Geometry and $W_\infty$ Strings } 

\bigskip

To finalize this work we shall concentrate in the reasons why we believe that $W_\infty$ symmetry,  and its
higher dimensional extensions,  volume-preserving diffs,   
are very important in $M$ theory. In particular, we will perform a Vasiliev star product deformation
of the BF-CS-Higgs inspired action in {\bf 4.1} which generated the MMCW actions for gravity plus the  cosmological constant and 
Gauss-Bonnet invariant.  In the last sections we have obtained $AdS_4$ as a vacuum solution of the  BF-CS-Higgs inspired model 
and argued why the large $N$ limit of $SU(N)$ YM in $curved$  spaces can be subsumed into a higher dimensional 
gravitational theory . 

The main points of this section are : 

( i ) To realize that noncritical $W_\infty$ strings , moving in $AdS_4 \times S^7$ , are the sought-after theories 
with higher $conformal$  spin symmetries of an effective $3D$  " worldsheet "   $\leftrightarrow  $  
Vasiliev's higher spin gauge symmetries of the target $AdS_4$ spacetime $\leftrightarrow 3D$ higher conformal
spin symmetries associated with the conformal group of the boundary $SO(3, 2)$ , the isometry group of $AdS_4$.

( ii ) The effective $3D$ higher conformal spin field theory of $W_\infty$ strings 
is related to  the $3D$ continuous  $SU(\infty) ( SL(\infty))  $ Toda theory obtained from a Killing symmetry reduction of 
$4D$ Self Dual Gravity [22, 36] .  And the latter can be obtained from the $SU(\infty)$ SDYM equations in four-dimensions [ 22 ] . 

(iii)  Such $3D$ continous Toda theory can also be obtained from the equations of motion for a Self Dual Membrane [22] . Membranes 
live on the boundary of $AdS_4$ ,  hence , one should expect a noncritical $W_\infty$ string sector of the 
membrane's spectrum to live in $AdS_4\times S^7  
\leftrightarrow $  Vasiliev's higher spin fields propagating in the target  $AdS_4$ spacetime. 

(iv) The alledged $ D = 11$ critical dimension of the supermembrane coincides precisely with the value of the dimension 
associated with a $noncritical $ $W_\infty$ superstring devoid of quantum superconformal anomalies [ 22 ] . 
Using a BRST analysis, we have shown that a nilpotent BRST charge operator, associated with the
noncritical $W_\infty$ superstring , can be constructed by adjoining a 
$ q = N+1$ unitary superconformal model of the  super $W_N$ algebra , 
in the $ N = \infty$ limit,  to a critical $W_\infty$ superstring spectrum.  
Therefore, we have an anomaly-free $noncritical$  $W_\infty$ superstring moving in $ D = 11$.  

Similar analyis followed for the bosonic $noncritical$  $W_\infty$ string and we found [22] that 
$ D = 27$  was the required dimension of the target spacetime . This is also the dimension of the alledgedly anomaly-free bosonic membrane.

(v) Conformal invariance was instrumental in contraining the types of backgrounds that strings can propagate. 
Conformal invariance at the quantum level demanded the vanishing of the $\beta$ functions associated with the worldsheet couplings. 
And this implied that the target spacetime backgrounds must be classical solutions of the ( super ) gravitational equations of motion : 
like  $ AdS_4 \times S^7$. 

Extending these arguments to $W_\infty$ strings , propagating on $AdS_4 \times S^7$ ,  
an extended anomaly-free quantum CFT based on the $W_\infty$ symmetry algebra ( area-preserving ) 
of the $W_\infty$ string , should impose  strong constraints 
on the type of target spacetime backgrounds they can propagate , and hence $AdS_4 \times S^7$ should be one of them  
$\leftrightarrow $  a  ( super ) membrane living at the boundary of $AdS_4$.  

In one scoop we can argue that :  
This  particular noncritical $W_\infty$  (  super ) string theory   behaves  like  
a Self Dual Supermembrane  moving in target $ 11D $ backgrounds  
Based on the correspondence between Self Duality and Conformal invariance [36] , it is  reasonable to assume 
that this particular noncritical $W_\infty$ string theory  actually lives on the $boundary$ of $AdS_4$. In the parlance of 
CFT, one is " gluing" the CFT associated with the noncritical $W_\infty$ string ( an effective $3D$ theory ) onto the conformal 
boundary of $AdS_4$.

The other reasons which support our belief that $W$ symmetry is of importance in $M$ theory are the following :

$\bullet$ $2D~W_\infty$ geometry is related  to the 
Fedosov Deformation program associated with the symplectic
geometry of the cotangent bundles  of $2D$ Riemannian surfaces; the crucial  role that  
$4D$ Self Dual Gravity has in the construction of $w_\infty$ gravity actions was emphasized earlier by Hull [18]. 
Geometric induced actions for $W_\infty$ gravity based on the coadjoint
orbit method associated with $SL(\infty,R)$ WZNW models
were constructed by Nissimov, Pacheva and Vaysburd [23].  See also the work by Chapline and Rodgers [44, 70 ] . 
What this suggests is that a restricted higher-dim gravitational theory is connected to $w_\infty$ gravity in lower dimensions.

$$ 4D~ Self ~ Dual~ Gravity \leftrightarrow 2D~ w_\infty~ gravity $$

$\bullet$  $W_\infty$ gravity , in the lightcone gauge, has a hidden $SL(\infty,R)$ Kac-Moody symmetry.  Likewise, the
$SL(\infty)$ Toda model obtained from a rotational Killing symmetry
reduction of $4D$ Self Dual Gravity  ( an effective  $3D$ theory) has a $w_\infty$
symmetry [36 ].  $SL(\infty)$ Toda models can also be constructed via a Drinfeld-Sokolov Hamiltonian 
reduction of constrained WZNW models . 
Once again we can see the intricate relationship between self duality and conformal field theory in higher dimensions.

$\bullet$  What perhaps is the most significant and salient feature of Chern-Simons $p$-branes is 
the fact that they admit an infinite number of secondary constraints 
which form an infinite dimensional closed algebra with respect to the Poisson bracket. [11] 
Such algebra $contains$ the clasical $w_{1+\infty}$ as a subalgebra . The latter algebra corresponds to the 
area-preserving diffeomorphisms of a cylinder.; 
the $w_{\infty}$ algebra corresponds to the area-preserving diffs of  a plane; $su(\infty)$ 
for a sphere.....[14]. 
These $w_{\infty}$ algebras are the higher conformal spin $s=2,3,4.....\infty$ 
algebraic extensions of the $2D$ Virasoro algebra. The topology of the two-dim surface determines the type of 
infinite-dim algebras involved.

$\bullet$  Deformation quantization, nonlocality and higher spin algebras.  

A ( Moyal ) Quantization $deforms $ the $w_\infty$ to $W_\infty $ [ 58-60    ]. The latter algebras admit central extensions in all 
higher conformal spin sectors. The former only admitted central charges in the Virasoro sector.  
Higher Spin Algebras ( superalgebras) in dimensions greater than two
have been described by   Vasiliev and in [15] were used to cosntruct 
higher spin gauge interactions of massive particles in $AdS_3$ spaces.
These higher spin algebras have been instrumental lately in [16] to
construct  the   $N=8$ Higher Spin Supergravity in $AdS_4$ which is
conjectured to be the field theory limit  of M theory  on $AdS_4
\times S^7$.

Crucial in the construction of the Vasiliev higher spin algebras is
the noncommutative star products and the fact that these algebras required an
Anti de Sitter space. For the relevance of Moyal Brackets in $M$
theory we refer to Fairlie [19].  
It has been speculated that the $W_\infty$-symmetry of $W_\infty$
strings after  a Higgs-like
spontaneous symmetry breakdown yields the infinite massive tower of
string states.

Moyal star products are non-local due to the infinite number of
derivatives. This nonlocality in phase space is translated into a spacetime nonlocality. 
This in conjunction with the fact that Anti de Sitter spaces
are required may be relevant in understanding  the $AdS/CFT$
duality conjecture.  
$W $ algebras were essential to identify the missing states in the $AdS/CFT$ correspondence [ 45] . 
Higher Derivative Gravity is also very relevant in the AdS/CFT correspondence [49] .

$\bullet$ Singletons/doubletons and compositeness . 

The  massless excitations of the $CFT$ living on
the projective boundary of  
Anti de Sitter space, associated with the propagation of branes on $AdS_d \times S^{D-d}$,   are 
composites of singleton, doubletons...fields [17]. 
The $SO(5)$ sigma models actions (8c) and the corresponding $SO(4)YM$ fields in
eqs-(7a,7b) are based on $composite~ fields$  ; i.e made out of the $n^i
(x)$. New actions for all $p$-branes where  the $analogs$ of $S$
and $T$ duality symmetries were  built in, already from the start, were given in
[20] based on the composite antisymmetric tensor field theories of the
volume-preserving diffs group of Guendelman, Nissimov ,
Pacheva and the local field theory reformulation of extended objects
given by
Aurilia, Spallucci and Smailagic [21].  
This supports the idea that compositeness may be a crucial ingredient
in the formulation of $M$ theory.

Having described our reasons why $W$ algebras seem to be of  essential importance for $M$ theory , 
we will construct the higher spin version of the 
BF-CS-Higgs action that generated $AdS_4$ as a natural vacuum solution in {\bf 4.1}.

Vasiliev's construction of higher spin gauge theories and their couplings to higher spin matter currents on $AdS$ spaces can 
be attained by introducing a suitable 
noncommutative but associative Vasiliev star product on an 
auxiliary ( commuting ) Grassmaniann even " fermionic phase space " 
whose deformation parameter is the 
inverse length scale characterizing the size of $AdS_4$'s throat $\lambda =  r^{-1} $.

The Vasiliev star product encoding the nonlinear and nonlocal 
higher spin fields dynamics is defined taking advantage of the local isomorphism $ so(3,2 ) \sim sp ( 4, R  ) $. It has the same form 
as the Baker integral representation of the star product :

$$ ( F* G) ( Z, Y ) =( { 1\over 2 \pi } )^{4} 
\int d^2 u ~d^2 {\bar u }~ d^2 v ~d^2 {\bar v} ~ 
e^{ i (u^\alpha v_\alpha - {\bar u}^{\dot \alpha} {\bar v}_{\dot \alpha}) } 
 ~F ( Z+U, Y+U     ) G(  Z- V, Y +V    ) . \eqno (82  ) $$
where the spinorial coordinates :

$$ Z^m = ( z^\alpha, {\bar z}^{\dot \alpha}) .~~~ Y_m = (y_\alpha, {\bar y}_{\dot \alpha}). 
~~~ \alpha, \beta = 1, 2. ~~~{\dot \alpha} = 1, 2. $$
yield a particular realization of the Weyl algebra : 

$$ [ y_\alpha, y_\beta ]_* = y_\alpha * y_\beta - y_\beta* y_\alpha = - [ z_\alpha, z_\beta ]_*  = 
2 i C_{\alpha \beta } . ~~~
[ y_\alpha, z_\beta ]_* = 0. ....$$

We must emphasize that one must $not$  confuse Vasiliev's defomation of 
$SO(3, 2)$ with the Moyal 
deformation of $SU(N)$.   
One can introduce internal $SU(N)$ gauge symmetries via the Chan-Paton mechanism at the 
boundary of open strings. This is tantamount of introducing $w_\infty$ algebras with internal $SU(N)$ symmetries [ 53  ]. 
The large $N$ ( infinite )  colour limit  of such algebras have been coined 
$w_\infty^\infty$ in [ 61  ] and correspond to symplectic difeomorphisms in $ D = 4 $.

The Vasiliev-algebra valued one-form   $W =  dx^\mu W_\mu$  contains the  master field $W_\mu $ 
that generates after a Taylor expansion 
all the higher massless spin gauge fields : 

$$ W =  dx^\mu W_\mu  = dx^\mu W_\mu ( x| Z, Y, Q ) . ~~~ $$

$$ W_\mu ( x| Z, Y, Q ) = \sum  W_{\mu,  \alpha_1 \alpha_2....\beta_1 \beta_2....} (x|Q) 
z^{\alpha_1}z^{\alpha_2}.....y^{\beta_1} y^{\beta_2}....                        \eqno ( 83) $$
where $ Q$ is a  discrete set of  Clifford variables ( Klein operators ) that anticommute with all spinorial coordinates. 

The Vasiliev-algebra-valued field strengths $ F(W)$ are : 

$$ F ( W ) = dW (x| Z, Y, Q )  +  W ( x| Z, Y, Q )  * \wedge ~ W (x|Z, Y, Q ) . \eqno ( 84 ) $$

 The matter fields belong to the  Vasiliev-algebra valued $ 0$-forms : $ \Phi ( x|Z, Y, Q ) $ 
 and are the generalizations of the original Higgs scalars $\phi^A$  in the action (44) .  
 There are Stueckelberg compensating spinor-valued fields 
 $ S_\alpha ( x| Z, Y, Q ) $ that carry only pure gauge degrees of freedom and act as covariant differentials along the 
 fermionic $Z$ directions. There are auxiliary fields as well in order to have 
$off$-shell  realizations of Vasiliev star algebra $SO(3, 2 )^* $ .   
 Auxiliary fields are required also to have off-shell realizations of Supersymmery. 
The precise form of the auxiliary fields for these off-shell realizations of the 
higher spin algebras is unknown at the moment and 
lies beyond the scope of this work. 

Nevertheless, we shall go ahead and just write down the Vasiliev star deformation of the orginal 
BF-CS-Higgs  inspired action (44  ) that generated  gravity with a cosmological constant , after a symmetry breaking :

 $$ S^*_{BF-CS} = \int dY ~dZ ~ [ \Phi *\wedge~  F(W) * \wedge~ F(W)  + \Phi *\wedge~  d\Phi * \wedge ~ d\Phi * \wedge ~  
d\Phi * \wedge  ~d\Phi ]  . \eqno ( 85 ) $$
the deformed Higgs potential is : 
$$ V_H^* ( \Phi ) = ( \Phi * \Phi - v ) * ( \Phi * \Phi - v ) . \eqno ( 86) $$
where one must add the contribution from the Stueckelberg and auxiliary fields as well and some over 
the discrete Clifford variables. 
The integration w.r.t the auxiliary spinorial variables plays a similar role 
of performing the internal " Trace " operation in ordinary YM theories. 
The integrand is a spacetime four-form . 
An integration of the above action w.r.t the auxiliary fermionic variables will yield an 
$effective$ action in $AdS_4$ that encodes the 
 nonlinear dynamics of the infinite number of higher spin massless gauge fields .  
As previously said, Gauge theories based on the Virasoro and $w_\infty$ algebras 
 have been constructed in [37-39 ] . A similar integration w.r.t the internal degrees of freedom reproduces an 
spacetime effective action involving all the infinite modes associated with the Virasoro and $w_\infty$ gauge fields [37-39]

 A BF-CS-Higgs-YM based on $SU(N)$ can be constructed as well since , an $SU(N)$  internal symmetry  
 can be introduced via the Chan-Paton mechanism at the end-points of open $W_\infty$ strings. 
The Vasiliev fields are now  
$matrix$-valued : $ W^{ij} ( x| Z, Y, Q ) $. 
The large $N$ limit can be attained by the techniques of the previous sections leaving  
an effective $12$-dim field theory ( ignoring the discrete degrees of freedom associated with the Klein variables ) : 
$$ W ( x| Z, Y, Q| q^i, p^i ) .\eqno ( 87) $$

There are $4$ coordinates $ x^\mu$. The auxiliary fermionic variables amount to $ 4$ 
degrees of freedom ( fermions carry $1/2 $ of a boson ). 
There are $4$ degrees of freedom associated with $ (q^1, q^2, p^1, p^2 )$ obtained from the 
star products deformation of $SU(N)_*$.  We  find it very encouraging that an effective 
$ D = 12 $  theory emerges very naturally from the large $N$ limit of the 
Vasiliev matrix-valued algebra fields $ W^{ij} $ living in $ AdS_4$.  $ D = 12$ is the realm of Vafa's $F$ theory and Bars $S$ theory.

We stop at this point. A lot remains to be done.  We close this work with a few interesting thoughts. The $n\rightarrow
\infty$ limit of the $SO(2n,1)~\sigma$-models  studied in {\bf 2} is connected with the
$D=2n\rightarrow \infty$ limit of the Euclidean signature $AdS_{2n}$ space. Interestingly
enough, Zaikov has pointed out that in the $D=\infty$ limit,  
Chern-Simons $p$-branes acquire true {\bf local}  dynamics ! 
Infinite Dimensions based on a Hierarchy of infinite nested spaces 
will give us a unique vantage point   : 
one master gauge field in infinite dimensions $\leftrightarrow $  infinite higher spin massless fields in lower dimensions.  

Since  Zaikov's Chern-Simons branes {\bf are}  High dimensional Knots , 
its relation to algebraic {\bf K, L} theory has to be explored 
deeper [  47 ] . The fact that an Effective $ D = 12 $ theory emerged from the large $N$ limit of 
Vasiliev's  matrix valued higher spin algebras , in connection to introducing $SU(\infty)$ Chan-Paton factors to the end-points of 
open $W_\infty$ strings ,  warrants investigation. The infinite colour limit is given by algebra of symplectic diffs in four dimensions 
$w_\infty^\infty$ [ 61 ].

\bigskip
\centerline{\bf  Acknowledgements}
\bigskip
I wish to thank Euro Spallucci, Stefano Ansoldi, Antonio Aurilia, Alex Granik  and George Chapline for many frutiful discussions at the
early stages of this work; to
M. Peskin for his assistance at SLAC, Stanford ; to  E. Abdalla,
M. Gomes, M. Sampaio ,
E. Valadar for their help at the Instute of Physics, Sao Paulo and the
UFMG, Belo Horizonte.  
Special thanks goes to M. L
Soares de Castro and M. Fernandes de Castro
for their warm hospitality in Belo Horizonte, Brazil where this work was initiated. 
\bigskip
\centerline{\bf References}
\bigskip

1. B. Dolan, D.H Tchrakian : Phys. Letts {\bf B 198} (4) (1987) 447. 

2. B. Dolan, Tchrakian : Phys. Letts {\bf B 202} (2) (1988) 
211.

3.B. Feisager, J.M Leinass : 
Phys. Letts {\bf B 94}  (1980) 192.

4. C. Castro : "Remarks on  Spinning Membrane Actions  "hep-th/0007031

5. Lindstrom, Rocek  : Phys. Letts {\bf B 218}  (1988) 
207.

6. J. Maldacena : Adv. Theor. Math. Phys {\bf 2} (1998) 231. hep-th/9711200.

J.L Petersen : "Introduction to the Maldacena Conjecture "hep-th/9902131. 

7. N. Berkovits, C. Vafa , E. Witten : "Conformal Field Theory of $AdS$ Backgound 

with Ramond-Ramond Flux" hep-th/9902098.

8. J. de Boer, S.L Shatashvili : "Two-dimensional Conformal Field Theory 

on $AdS_{2n+1}$ Backgrounds ". hep-th/9905032. 

9. A. Belavin, A.M Polyakov, A.S Schwarz and Y. S. Tyupkin : 

Phys. Letts {\bf B 59}  (1975) 85.

10.D.H Tchrakian : Jour. Math. Phys. {\bf 21} (1980) 166

11 R.P Zaikov : Phys. Letts {\bf B 266}  (1991) 303. Phys. Letts {\bf B 263}  (1991) 209. 

R.P Zaikov : Phys. Letts {\bf B 211}  (1988) 281.  

R.P Zaikov : "Chern-Simons $p$-Branes and $p$-dimensional Classical $W$ Algebras ". 

hep-th/9304075.

12. Y. Ne'eman, E. Eizenberg : "Membranes and Other Extendons ( $p$-branes) 

"World Scientific Lecture Notes in Physics vol. {\bf 39} 1995. 

J. Hoppe : Ph.D Thesis MIT (1982).

13. E. Witten : Comm. Math. Phys. {\bf 121} (1989) 351. 

Nucl. Phys. {\bf B 322} (1989) 629.

14. P. Bouwknegt, K. Schouetens : "$W$-symmetry in Conformal Field Theory "

Phys. Reports {\bf 223} (1993) 183-276.

15. M. Vasiliev : " Higher Spin Gauge Theories, Star Product and AdS spaces " 

hep-th/9910096

M. Vasiliev , S. Prokushkin : "$3D$ Higher-Spin Gauge Theories with Matter". 

hep-th/9812242, hep-th/9806236. 

16.E Sezgin, P. Sundell : "Higher Spin $N=8$ Supergravity in $AdS_4$". 

hep-th/9805125; hep-th/9903020. 

17. M. Duff : "Anti de Sitter Spaces, Branes, Singletons, Superconformal 

Field Theories and All That "hep-th 9808100. 

S. Ferrara, A. Zaffaroni : "Bulk Gauge Fields in $AdS$ Supergravity 

and Supersingletons "hep-th/9807090. 

M. Flato, C. Fronsdal : Letts. Math. Phys. {\bf 44} (1998) 249. 

M. Gunaydin. D. Minic, M. Zagermann : "$4D$ Doubleton Conformal Field Theories, CPT and $IIB$ String 

on $AdS_5 \times S^5$. hep-th/9806042.

18. B. Fedosov : " Jou. Diff. Geometry {\bf 40} ( 1994) 213.

C. Castro : "W-Geometry from Fedosov Deformation Quantization " 

Jour. Geometry and Physics {\bf 33} ( 2000 ) 173.

19. D. Fairlie : "Moyal Brackets in M theory "hep-th/9707190. 

Mod. Phys. Letts {\bf A 13 } (1998) 263 

20. C. Castro : Int. Jour. Mod. Phys. {\bf A 13 } (6) (1998) 1263.

21.E.I  Guendelman, E. Nissimov, S. Pacheva : 

"Volume-Preerving Diffs versus Local gauge Symmetry : hep-th/9505128.

H. Aratyn, E. Nissimov, S. Pacheva  : Phys. Lett {\bf B 255} (1991) 359.

A. Aurilia, A. Smailagic, E. Spallucci : Phys. Rev. {\bf D 47} (1993) 2536.

22. C. Castro : Jour. Chaos, Solitons and Fractals {\bf 7} (5) (1996) 711.

C. Castro : " Jour. Math. Phys. {\bf 34} ( 1993) 681.  

C. Castro : Phys Lets {\bf B 288  }  (1992) 291.

23.E Nissimov, S. Pacheva , I. Vaysburd : "$W_\infty$ Gravity, a Geometric Approach ". 

hep-th/9207048. 

24. G. Chapline : Jour. Chaos, Solitons and Fractals  {\bf 10} (2-3) (1999) 311.

 Mod. Phys. Lett {\bf A 7} (1992) 1959. Mod. Phys. Lett {\bf A 5} (1990) 2165. 

25. C. Vafa : "Evidence for F Theory"hep-th/9602022

26. I. Bars : ``Two times in Physics.'' hep-th/9809034.

27. P. Horava : "$M$ Theory as a Holographic Field Theory "hep-th/9712130. 

28. L. Smolin  : "Chern-Simons theory in $11$ dimensions 

as a non-perturbative phase of $M$ theory ".

29. M. Gomes, Y. K Ha : Physics Letts {\bf 145 B} (1984) 235. 

Phys. Rev. Lett {\bf 58} (23) (1987) 2390.

30. P. Tran-Ngoc-Bich : Private Communication.  

31. H. Garcia-Compean, J. Plebanski, M. Przanowski :.hep-th/9702046.
 
``Geometry associated with SDYM and chiral approaches to Self Dual Gravity ``. .

32. A. Connes, M. Douglas, A. Schwarz : ``Noncommutative Geometry and

Matrix Theory : Compactification on Tori ``hep-th/9711162.

33- S. Ansoldi, C. Castro, E. Spallucci : Phys. Lett {\bf B 504} (2001) 174. 

34- S. Ansoldi, C. Castro, E. Spallucci : Class. Quant. Grav. {\bf 18} (2001) L 23. 

C. Castro : " Branes from Moyal Deformation quantization of GYM " hep-th/9908115. 

35- S. Ansoldi, C. Castro, E. Spallucci : " A QCD Membrane " 

to appear in Class. Quan. Gravity. hep-th/0106028

36-Q.H. Park : Int. Jour. Mod. Phys {\bf A 7} ( 1992) 1415.

37-Y.Cho, K, Soh, Q. Park,  J. Yoon : Phys. Lets {\bf B 286} ( 1992) 251.  :

38-Y. Cho, W. Zoh  : Phys. Review {\bf D 46} ( 1992) 3483.  

39-W. Zhao : Jour. Math. Phys. {\bf 40} ( 1999 ) 4325. 

40-L. Freidel, K. Krasnov :  " $BF$ deformation of Higher dimensional Gravity " 

hep-th/9901069

41-F. Wilcezk : Physical Review Letts {\bf 80} ( 1998) 4851.

42- E. Sezgin : " Area-preserving Diffs, $w_\infty$ algebras , $w_\infty$ gravity " 

hep-th/9202080. 

E. Sezgin : " Aspects of $ W_\infty$ Symmetry " hep-th/9112025 .  

43- C. Hull : Phys. Letts {\bf B 269 } ( 1991) 257.

44-Rodgers :  " BF ....." hep-th/9203067

45-S. Korden : " W algebras from the AdS/CFT correspondence " hep-th/0011292

46- N. Ikeda : " Deformation of BF theories, Topological Open Membranes ...." 

hep-th/0105286.

47-A. Ranicki : " High Dimensional Knots , Algebraic Surgery in Codimension Two " 
Springer Verlag , 1998. 

48-A.Cattaneo, P. Cotta-Ramousino, F. Fulcito, M. Martellino, M. Rinaldi, 

A. Tanzini, M. Zeni : " Comm. Math. Phys. {\bf 197} ( 1998) 571 .

49- M. Fukuuma, S. Matsuura, T. Sakai : " Higher Derivative Gravity  

and the AdS/CFT correspondence " hep-th/0103187

50- T. Deroli , A. Verciu :  Jour. Math. Phys {\bf 38 } ( 11 ) ( 1997 ) 5515.  

E. Gozzi, M. Reuter :" Quantum Deformed Geometry of Phase Space " DESY-92-191.

51-E. Berghshoeff, A.Salam, E. Sezgin, Y. Tanii : Nuc. Phys. {\bf B 305 } (1988) 497.

52- H. Nicolai, E. Sezgin, Y. Tanii : Nuc. Phys. {\bf B 305 } (1988) 485 .

53- S. Odake, T. Sano : Phys. Letts {\bf B 258 } ( 1991) 369. 

54-S. W. MacDowell, F. Mansouri : Phy. Rev. Let {\bf 38} ( 1977) 739.

55- A. Chamseddine, P. West : Nuc. Phys. {\bf B 129} ( 1977) 39.

56-E. Guendelmann

57- C. Castro, E. Guendelmann, E. Spallucci : To appear.

58-I. Bakas. B. Khesin. E. Kiritsis : Comm. Math . Phy {\bf 151} (1993 ) 233.

59-K. Dasgupta, Z. Yin : " Nonabelian Geometry " hep-th/0011034

60- C. Pope, L. Romans, X. Shen : Nuc. Phys, {\bf B 339} ( 1990) 191.

61-I. Bakas, E. Kiritsis : " Comon Trends in Mathematics and Physics " 

Eds. T. Eguchi, ....Prog. Theor. Physics Suppl. ( 1990 )

62- I. Oda : Chaos, Solitons and Fractals {\bf 10} ( 2-3 ) ( 1999) 423. 

63- G. Dito, M. Flato, D. Sternheimer, L. Takhtajan : 

" The deformation quantization of Nambu-Poisson mechanics " hep-th/9602016 

H. Awata, M. Li, G. Minic, T. Yoneya : " Quantizing the Nambu Poisson brackets " 

hep-th/9906248 . 

G. Minic : " M theory and Deformation Quantization " hep-th/9909022 

64-M. Kontsevich : " Deformation quantization of Poisson manifolds " q-alg/9709040

65-A. S. Cattaneo, G. Felder : math.QA/9902090.

66-N. Seiberg, E. Witten : JHEP 9909 ( 032 ) ( 1999) hep-th/9908142 

67-J. Baez : Lectures Notes in Physics, Springer {\bf 543} ( 2000 ) 25-94. 

Class. Quan. Grav. {\bf 15} ( 1998) 1817. 

68-J. de Boer :  " The Holographic Renormalization Group "  hep-th/0101026.

69- Y. M Cho , H. Lee, D. Park : " Faddeev-Niemi Conjecture and 

the Effective Action for QCD " hep-th/0105198. 

W. Bae, Y. M. Cho, S. Kimm : " QCD versus Skyreme-Faddeev Theory "      

hep-th/0105163 

70-  G. Chapline, K. Yamagishi : Class. Quan. Gravity {\bf 8 } ( 1991) 427. 

71- S. Ansoldi, A. Aurilia, C. Castro, E. Spallucci :  in preparation

\bye